\documentclass[aps, twocolumn, pra, superscriptaddress, amsmath, showpacs, tightenlines, letterpaper] {revtex4}
\usepackage{epsfig,graphicx,times}
\usepackage{amstext}
\usepackage{amsmath}            
\usepackage{amssymb}            
\usepackage{amsthm}
\usepackage{leftidx}
\usepackage{mathrsfs}
\usepackage{graphicx}           
\usepackage{subfigure}
\usepackage{latexsym}
\usepackage{bm}
\usepackage{epstopdf}
\usepackage{epsfig}
\usepackage{color}
\begin{document}
\title[Engineering arbitrary entanglement of two microwave resonators]{Generating nonclassical photon-states via longitudinal couplings between
superconducting qubits and microwave fields}
\author{Yan-Jun Zhao}
\affiliation{Institute of Microelectronics, Tsinghua University, Beijing 100084, China}
\author{Yu-Long Liu}
\affiliation{Institute of Microelectronics, Tsinghua University, Beijing 100084, China}
\author{Yu-xi Liu}
\email{yuxiliu@mail.tsinghua.edu.cn}
\affiliation{Institute of Microelectronics, Tsinghua University, Beijing 100084, China}
\affiliation{Tsinghua National Laboratory for Information Science and Technology (TNList),
Beijing 100084, China}
\affiliation{CEMS, RIKEN, Saitama 351-0198, Japan}
\author{Franco Nori}
\affiliation{CEMS, RIKEN, Saitama 351-0198, Japan}
\affiliation{Physics Department, The University of Michigan, Ann Arbor, Michigan
48109-1040, USA}


\begin{abstract}
Besides the conventional transverse couplings between\textbf{ }superconducting
qubits (SQs) and electromagnetic fields, there are additional longitudinal
couplings when the inversion symmetry of the potential energies of the SQs is
broken. We study nonclassical-state generation in a SQ which is driven by a
classical field and coupled to a single-mode microwave field. We find that the
classical field can induce transitions between two energy levels of the SQs,
which either generate or annihilate, in a controllable way, different photon
numbers of the cavity field. The effective Hamiltonians of these
classical-field-assisted multiphoton processes of the single-mode cavity field
are very similar to those for cold ions, confined to a coaxial RF-ion trap
and driven by a classical field. We show that arbitrary\textbf{ }%
superpositions of Fock states can be more efficiently generated using these
controllable multiphoton transitions, in contrast to the single-photon
resonant transition when there is only a SQ-field transverse coupling. The
experimental feasibility for different SQs is also discussed.

\end{abstract}
\maketitle

\pagenumbering{arabic}

\section{Introduction}

Superconducting qubit\textbf{ }(SQ) circuits~\cite{r1,r2,r3,r4,r5,r6,r7,r8}
possess discrete energy levels and can behave as artificial \textquotedblleft
atoms". In contrast to natural atoms, with a well-defined inversion
symmetry of the potential energy, these artificial atoms can be controlled by
externally-applied parameters (e.g., voltage or magnetic
fluxes)~\cite{r1,r2,r3,r4,r5} and thus the potential energies for these qubits
can be tuned or changed from a well-defined inversion symmetry to a broken one.
Artificial atoms with broken symmetry have some new features which do not
exist in natural atoms. For example, phase qubits do not have an optimal
point~\cite{martinis,ustinov}, so for these the inversion symmetry is always broken.

When the inversion symmetry of these artificial atoms is broken, then the
selection rules do not
apply~\cite{liu2005,liu2010,Savasta2014_1,Savasta2014_2}, and
microwave-induced transitions between any two energy levels in multi-level SQ
circuits are possible. Thus, multi-photon and single-photon processes
(or many different photon processes) can coexist for such artificial
multi-level systems~\cite{liu2005,liu2010,naturephysics2008}.
Two-level natural atoms have only a transverse coupling between these two
levels and electromagnetic fields. However, it has been shown~\cite{liu2010} that there are
both transverse and longitudinal couplings between SQs and applied magnetic
fields when the inversion symmetry of the potential energy of the SQ is
broken. Therefore, the Jaynes-Cumming model is not suitable to
describe the SQ-field interaction when the inversion symmetry is broken.

Recently, studies of SQ circuits have achieved significant progress. The
interaction between SQ\textbf{ }circuits and the electromagnetic field makes
it possible to conduct experiments of quantum optics and atomic physics on a
chip. For instance, dressed SQ states (e.g., in
Refs.~\cite{liudressed,greenberg}) have been experimentally
demonstrated~\cite{large-charge3,wallraff2009}. Electromagnetically-induced
transparency (e.g.,
Refs.~\cite{orlando2004,goan,ian,falci,blais2010,HuiChenEIT,PengBoEIT}) in
superconducitng systems has also been theoretically studied. Moreover,
Autler-Townes splitting~\cite{PengBoEIT,atsET,atsEP,atsEF,Bsanders,atsET-2,atsET-3} and
coherent population trapping~\cite{CPT} have been experimentally demonstrated
in different types of SQs with three energy levels. Experiments have shown
that SQs can be cooled (e.g., Refs.~\cite{youjqprl,orlandon,Nori2008,cooling})
using similar techniques as for cooling atoms. Moreover, sideband excitations~\cite{side1,blais2007} have been observed experimentally~\cite{side2,side3} using superconducting circuits. Thus, SQs can be manipulated as trapped ions (e.g., in Ref.~\cite{trappedions,Liu2007,wei}), but
compared to trapped ions, the \textquotedblleft vibration mode" for SQs is
provided by an LC circuit or a cavity field.

In trapped ions~\cite{trappedions,Liu2007,wei}, multi-phonon transitions can be
realized with a laser field. Multi-photon processes in SQs with driving
fields~\cite{large2} have been experimentally observed (e.g., in
Refs.~\cite{NoriPR2010,multi1,multi2,large1,berns,large3}) when the inversion
symmetry is broken. Thus, here we will show how nonclassical photon states
can be generated, via multi-photon transitions of a single-mode electromagnetic
field in a driven SQ, when the longitudinal coupling field is introduced. We
will derive an effective Hamiltonian which is similar to the one for
trapped ions. The single-mode quantized field can be provided by
either a transmission line resonator (e.g.,
Refs.~\cite{youcavity,youcavity1,blais2004,walraff}) or an LC circuit (e.g.,
Refs.~\cite{mooij1,cooling}), where the SQ and the single-mode field have
both transverse and longitudinal couplings. In contrast to the generation of
non-classical photon states using a SQ inside a
microcavity~\cite{liuepl,liupra,martinis1,martinis2} with only a single-photon transition, we will show that the Hamiltonian derived here can be used to more efficiently
produce nonclassical photon states of the microwave cavity field when longitudinal-coupling-induced multiphoton transitions are employed.

Our paper is organized as follows. In Sec.~II, we derive an effective
Hamiltonian which is similar to the one for trapped ions. We also describe the analogies and
differences between these two types of Hamiltonians. In Sec.~III, we show how to engineer nonclassical
photon states using the multi-photon coupling between the driven SQ and the
quantized field. In Sec.~VI, we discuss possible experimental\textbf{
\color{red}}implementations \textbf{\color{black} }of these proposals for
different types of SQs. Finally, we present some discussions and a summary.

\begin{figure}[ptb]
\includegraphics[width=0.46\textwidth, clip]{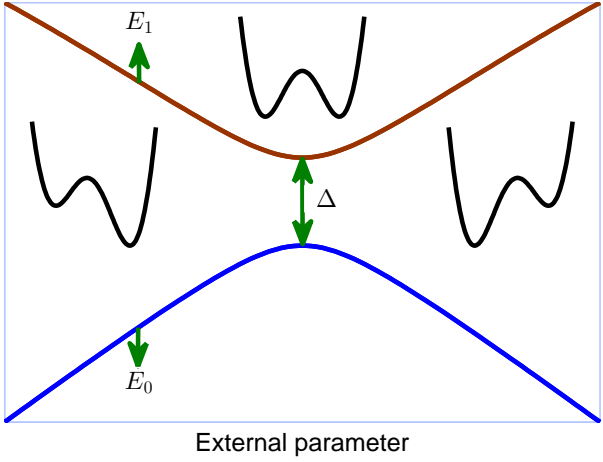}
\caption{(Color online) Schematic diagram showing how two energy levels change
with the external parameter for superconducting qubits. Here $E_{0}$ and
$E_{1}$ are the eigenvalues of the ground and excited states, respectively.
These vary with external parameters. For
charge and flux qubits, the external parameters are the electric voltage and
magnetic flux, respectively\color{black}. At the degenerate (or optimal)
point, where the external parameter takes a particular value, the energy
splitting reaches a minimum, $\Delta=\hbar\omega_{x}$, where
the double potential well is symmetric. In this case, there is only a SQ-field
transverse coupling. However, when the external parameter deviates from this
point, the double potential well is asymmetric, and there are both transverse
and longitudinal couplings between the SQs and the applied electromagnetic
field. }%
\label{fig1}%
\end{figure}

\section{Multi-photon process induced by a longitudinal coupling}

\subsection{Theoretical model}

As schematically shown in Fig.~\ref{fig1}, the shape of the potential energy
for some kinds of SQs (e.g., charge and flux qubits) can be adjusted (from symmetric to asymmetric and vice versa) by an external parameter, and thus the two energy
levels of SQs can also be controlled. For charge and flux qubits, the external
parameters are the voltage and the magnetic flux, respectively. However the
potential energy of the phase qubits is always broken, no matter how the
external field is changed. The generic Hamiltonians for different
types\textbf{ }of SQs can be written as
\begin{equation}
H_{q}=\frac{\hbar}{2}\omega_{z}\sigma_{z}+\frac{\hbar}{2}\omega_{x}\sigma_{x}.
\label{qubit1}%
\end{equation}
As in experiments, we assume that both parameters $\omega_{z}$ and $\omega
_{x}$ can be controlled by external parameters. The parameter $\omega_{z}=0$
corresponds to the optimal point and well-defined inversion symmetry of the
potential energy of the SQs. However, both nonzero parameters $\omega_{z}$ and
$\omega_{x}$ correspond to a broken inversion symmetry of the potential energy
of the SQs. Below, we first provide a general discussion based on the qubit
Hamiltonian in Eq.~(\ref{qubit1}), and then we will specify our discussions to
different types of SQs. The discussion of their experimental feasibilities
will be presented after the general theory.

Let us now assume that a SQ is coupled to a single-mode cavity field and is
driven by a classical field, where the Hamiltonian of the driven
superconducting qubits is\textbf{ }
\begin{equation}
H=H_{q}+\hbar\omega a^{\dagger}a+\hbar g\sigma_{z}(a+a^{\dagger})+\hbar
\Omega_{d}\sigma_{z}\cos(\omega_{d}t+\phi_{d}). \label{eq:1}%
\end{equation}
Here, $a^{\dagger}(a)$ is the creation (annihilation) operator of a
single-mode cavity field with frequency $\omega$. The parameter $\Omega_{d}$
is the coupling constant between the SQ and the classical driving field with
frequency $\omega_{d}$. The parameter $g$ is the coupling constant between the
SQ and the single-mode cavity field. The parameter $\phi_{d}$ is the initial
phase of the classical driving field.

Equation~(\ref{eq:1}) shows that there are transverse and longitudinal
couplings between the SQs and the electromagnetic field. This can become clearer
if we rewrite the Hamiltonian in Eq.~(\ref{eq:1}) in the qubit basis, that
is,
\begin{align}
H  &  =\frac{\hbar}{2}\omega_{q}\widetilde{\sigma}_{z}+\hbar\omega a^{\dagger
}a+\hbar g_{z}\widetilde{\sigma}_{z}(a+a^{\dagger})+\hbar g_{x}%
\widetilde{\sigma}_{x}(a+a^{\dagger})\nonumber\\
&  +\hbar\Omega_{dz}\widetilde{\sigma}_{z}\cos(\omega_{d}t+\phi_{d}%
)+\hbar\Omega_{dx}\widetilde{\sigma}_{x}\cos(\omega_{d}t+\phi_{d}),
\label{eq:1-1}%
\end{align}
with four parameters $g_{z}=g\cos\theta$, $g_{x}=-g\sin\theta$, $\Omega
_{dz}=\Omega_d\cos\theta$, and $\Omega_{dx}=-\Omega_d\sin\theta$. Here, the
parameter $\theta$ is given by $\theta=\arctan(\omega_{x}/\omega_{x})$, and the qubit eigenfrequency is $\omega_{q}=\sqrt{\omega
_{x}^{2}+\omega_{z}^{2}}$.

The Hamiltonian in Eq.~(\ref{eq:1-1}) shows that the qubit has both transverse and
longitudinal couplings to the cavity (driving) fields with transverse
$g_{x}$ ($\Omega_{x}$) and longitudinal $g_{z}$ ($\Omega_{z}$) coupling
strengths. When both $\omega_{z}=0$ and $\Omega_{d}=0$, Eq.~(\ref{eq:1-1}) is
reduced to
\begin{equation}
\widetilde{H}=\frac{\hbar}{2}\omega_{q}\widetilde{\sigma}_{z}+\hbar
g\widetilde{\sigma}_{x}(a+a^{\dagger}) \label{eq:2-1}%
\end{equation}
which has only the transverse coupling between the SQ and the single-mode
field. If we further make the rotating-wave approximation, then
Eq.~(\ref{eq:2-1}) can be reduced to the Jaynes-Cumming model, which has been
extensively studied in quantum optics~\cite{scullybook}. That is, there is
only a single-photon transition process when the qubit is at the optimal
point. However, the transverse and the longitudinal couplings between the SQ
and the single-mode field coexist, when the inversion symmetry of the
potential energy is broken and $\omega_{z}$ is nonzero for the SQs. As shown
below, this coexistence can induce multi-photon transitions between energy
levels of SQs and make it easy to prepare arbitrary nonclassical states of the
cavity field.

Below, we assume that both $\omega_{z}$ and $\omega_{x}$ are nonzero. We also
assume that the SQ and the quantized field satisfy the large-detuning
condition, that is,%

\begin{equation}
\omega_{q}=\sqrt{\omega_{x}^{2}+\omega_{z}^{2}}\,\gg\omega.
\end{equation}

In this case, the SQ and the quantized field are nearly decoupled from each
other when the classical driving field is applied to the SQs.

\subsection{Multi-photon processes and sideband excitations}

Let us now study how multi-photon processes can be induced via a longitudinal
coupling by first applying a displacement operator
\begin{equation}
D\left(  \eta\frac{\sigma_{z}}{2}\right)  =\exp\left[  \eta\frac{\sigma_{z}%
}{2}(a^{\dagger}-a)\right]  , \label{eq:V}%
\end{equation}
to Eq.~(\ref{eq:1}) with
\begin{equation}
\eta=2g/\omega.
\end{equation}
Thus $\eta$ is the normalized qubit-cavity coupling. It is also known as the
Lamb-Dicke parameter. Hereafter, we denote the picture after the transformation
$D\left(  \eta\sigma_{z}/2\right)  $ as the displacement picture. In this
case, we have an effective Hamiltonian
\begin{align}
H_{\mathrm{eff}}\ =\  &  DHD^{\dagger}\ =\ \frac{\hbar}{2}\omega_{z}\sigma
_{z}+\hbar\omega a^{\dagger}a+\hbar\Omega_{d}\sigma_{z}\cos(\omega_{d}%
t+\phi_{d})\nonumber\\
&  +\frac{\hbar}{2}\omega_{x}\left\{  \sigma_{+}\exp\left[  \eta\left(
a^{\dagger}-a\right)  \right]  +\mathrm{h.c.}\right\}  . \label{eq:Heff}%
\end{align}
From Eq.~(\ref{eq:Heff}) with $\Omega_{d}=0$, we find that if $n\omega
=\omega_{z}$, then the multiphoton processes, induced by the longitudinal
coupling, can occur between two energy levels formed by the eigenstates of the
operators $\sigma_{z}$. However, such process is not well controlled.
Moreover, $\omega_{z}$ is usually not perfectly equal to $n\omega$, for
arbitrarily chosen $n$. These problems can be solved by applying a classical
driving field, in this case $\Omega_{d}\neq0$.

To understand how the classical field can assist the cavity field to realize
multi-photon processes in a controllable way, let us now apply another
time-dependent unitary transformation
\begin{equation}
U_{d}\left(  t\right)  =\exp\left[  \frac{i}{\hbar}H_{d}\left(  t\right)
\right]  \label{eq:Ud}%
\end{equation}
to Eq.~(\ref{eq:Heff}) with the Hamiltonian $H_{d}$ defined as
\begin{equation}
H_{d}\left(  t\right)  =\frac{\hbar\Omega_{d}\sigma_{z}}{\omega_{d}}%
\sin(\omega_{d}t+\phi_{d}), \label{eq:Hd}%
\end{equation}
and then we can obtain another effective Hamiltonian
\begin{align}
H_{\mathrm{eff}}^{\left(  d\right)  }  &  =U_{d}H_{\mathrm{eff}}U_{d}%
^{\dagger}-iU_{d}\frac{\partial U_{d}^{\dagger}}{\partial t}=\frac{\hbar}%
{2}\omega_{z}\sigma_{z}+\hbar\omega a^{\dagger}a\nonumber\\
&  +\frac{\hbar\omega_{x}}{2}\sum_{N=-\infty}^{\infty}\left\{  J_{N}\sigma
_{+}B_{N}(t)+\text{H.c.}\right\}  , \label{eq:2}%
\end{align}
where the time-dependent expression $B_{N}\left(  t\right)  $ is given as
\begin{equation}
B_{N}\left(  t\right)  =\exp\left[  \eta\left(  a^{\dagger}-a\right)
+iN(\omega_{d}t+\phi_{d})\right];
\end{equation}
$J_{N}\equiv J_{N}\left(  {x_{d}}\right)  $ is the $N$th Bessel function of
the first kind, with $x_{d}=2\Omega_{d}/\omega_{d}$ and $J_{N}({x_{d}%
})=(-1)^{N}J_{-N}({x_{d}})$, and $\eta$ is similar to the Lamb-Dicke parameter
in trapped ions ~\cite{trappedions,wei}. Via the unitary
\begin{equation}
V_{0}\left(  t\right)  =\exp\left[  \frac{i}{\hbar}H_{0}t\right]  ,
\label{eq:V0}%
\end{equation}
with
\begin{equation}
H_{0}=\frac{\hbar}{2}\omega_{z}\sigma_{z}+\hbar\omega a^{\dagger}a,
\end{equation}
we can further expand the Hamiltonian in Eq.~(\ref{eq:2}), in the interaction
picture, into
\begin{equation}
H_{\mathrm{int}}=\frac{\hbar}{2}\omega_{x}\sum_{N,m,n}\left\{  J_{N}%
^{mn}(t)\sigma_{+}a^{\dagger m}a^{n}+\mathrm{h.c.}\right\}  , \label{eq:3}%
\end{equation}
with%
\begin{align}
J_{N}^{mn}(t)  &  =\frac{(-1)^{n}J_{N}}{m!n!}\eta^{m+n}\exp\left[  -\frac
{1}{2}\left(  \eta\right)  ^{2}\right] \\
&  \times\exp\left[  iN(\omega_{d}t+\phi_{d})+i\left(  m-n\right)  \omega
t+i\omega_{z}t\right]  .\nonumber
\end{align}

Equation~(\ref{eq:3}) clearly shows that the couplings between the SQs and the
quantized cavity fields can be controlled via a classical field when they are
in the large-detuning regime. Comparing the Hamiltonian in Eq.~(\ref{eq:3})
with that for the trapped ions~\cite{trappedions,wei}, we find that the
Hamiltonian in Eq.~(\ref{eq:3}) is very similar to that of the two-level ion,
confined in a coaxial-resonator-driven rf trap which provides a harmonic
potential along the axes of the trap. Therefore, in analogy to the case of
trapped ions, there are two controllable multi-photon processes (called
\textit{red and blue sideband }excitations, respectively) and one
\textit{carrier process}:

\begin{description}
\item (i) when $n>m$\textbf{, }with $n-m=k$, and the transition satisfies the
resonant condition $N\omega_{d}=\omega_{z}-k\omega$\textbf{, }with
$N,\,k=1,\,2,\,3,\cdots$, the driving frequency $N\omega_{d}$ is red-detuned
from the qubit frequency $\omega_{z}$. Thus, we call this multi-photon process
 the \textit{red process.}

\item (ii) when $n<m$\textbf{, }with $m-n=k$, and the transition satisfies the
resonant condition $N\omega_{d}=\omega_{z}+k\omega$\textbf{, }with
$N,\,k=1,\,2,\,3,\cdots$, the driving frequency $N\omega_{d}$ is blue-detuned
from the qubit frequency $\omega_{z}$. Then we call this process the
\textit{blue} \textit{process}.

\item (iii) when $n=m$ and $\omega_{z}=N\omega_{d}$ ($N=1,\,2,\,\cdots$), the
driving field with $N$ photons can resonantly excite the qubit. We call this
transition the \textit{carrier process}.
\end{description}

However, there are also differences between the Hamiltonian for trapped
ions~\cite{trappedions,wei} and that in Eq.~(\ref{eq:3}). These differences are:

\begin{description}
\item (i) For a given frequency $\omega_{d}$ of the driving field, there is
only one  multi-photon-transition  process in the
system of trapped ions to satisfy the resonant condition, but the SQs can
possess several different multiphoton processes, resulting from the
longitudinal coupling between the classical field and the SQ. For instance,
with the given frequencies $\omega_{d}$ and $\omega$, and for the couplings
with the $N$th and $N^{\prime}$th Bessel functions, two transitions with the
red sideband resonant conditions: $N\omega_{d}=\omega_{z}-k\omega$ and
$N^{\prime}\omega_{d}=\omega_{z}-k^{\prime}\omega$\textbf{, }might be
satisfied. Once the condition $(N-N^{\prime})/(k^{\prime}-k)=\omega/\omega
_{d}$ is satisfied, then these two resonant transitions can simultaneously
occur. Similarly, for the case of blue-sideband
excitations, the condition that two resonant transitions
simultaneously occur for the $k$ and $k^{\prime}$ photon processes is
$(N-N^{\prime})/(k-k^{\prime})=\omega/\omega_{d}$. We can represent the
transition type in the sign of $k$ and $k^{\prime}.$ Thus if we want some
terms with $N^{\prime}$ unresonant, all we need to do is to let $(N-N^{\prime
})/(k-k^{\prime})\neq\omega/\omega_{d},$ i.e., $\omega_{z}\neq\omega\left(
N^{\prime}k-Nk^{\prime}\right)  /\left(  N^{\prime}-N\right)  .$ One
sufficient condition is that $\omega_{z}\neq\omega n/\left(  N^{\prime
}-N\right)  $ ($n=0,\pm1,\pm2,$ $\cdots$).

\item (ii) The Lamb-Dicke parameter $\eta$ for the trapped ions is determined
by the frequency of the vibration phonon, mass of the ion, and the wave vector
of the driving field. However the Lamb-Dicke parameter $\eta$ here is
determined by the frequency $\omega$ of the single-mode quantized field and
the coupling constant $g$ between the single-mode field and the SQ.

\item (iii) For multi-photon processes, the
coupling between trapped ions and the phonon is always on. However,
such processes can in principle be switched off at the
zeros of the Bessel functions of the first kind.

\item (iv) The term in Eq.~(\ref{eq:3}) with $N=0$ means that the driving
field has no help for the excitation of the SQ. Thus this term is neglected in
the following discussions. However, the driving field can always be used to
excite the trapped ions when certain resonant condition is satisfied.

\item (v) For trapped ions, the ratio between the transition frequency of the
qubit and the frequency of the vibration quanta is often about $10^{9}$. Thus
the upper bound for the photon number $k$ in the multiphotn process is about
$k=10^{9}$. However, in the SQ circuit, the frequency of the SQ can be several
tens of GHz, and the quantized cavity field can be in the regime of GHz. Thus
the photon number $k$ is not extremely large. For example\textbf{, } if
$\omega_{z}=20$ GHz and $\omega=2$ GHz, then the upper bound for $k$ is $10$.
\end{description}

\bigskip To compare similarities and differences, Table
~\ref{tab:tab1} lists the main parameters of the Hamiltonian for
trapped ions and those of the SQ in Eq.~(\ref{eq:2}). We should note that
the Lamb-Dicke parameter $\eta$ can become very large in circuit QED systems
in the ultrastrong~\cite{Niemczyk,Forn,Ultrastrong3} and
deep-strong~\cite{Casanova,Braak,Solano,Liberato} coupling regime. Our
discussion below is in the ultrastrong coupling, but can be straightforwardly
extended to the deep-strong coupling regime.

\begingroup\squeezetable\begin{table}[ptb]
\caption{Comparison of some parameters between the Hamiltonian in
Eq.~(\ref{eq:3}) and that of the trapped ions (e.g., in Ref.~\cite{wei}). Here LD refers to the Lamb-Dicke parameter.}%
\label{tab:tab1}
\begin{ruledtabular}
\begin{tabular}{c|p{2.5cm}|p{2.5cm}}
Parameters& Superconducting qubits (orders of magnitude)& Trapped ions (orders of magnitude) \\
\hline&&\\
LD parameters& $2g/\omega  \sim (0.2-1.8)$ & $\eta\sim (0.2-0.9)$ \\
\hline&&\\ Carrier Rabi frequencies & Renormalized $J_{N}\:\omega_x/2$ & Renormalized $\Omega$ \\
\hline&&\\ Driving field frequencies & $\omega_{d} \,(N=1,\,\cdots)$& $\omega_{L}$\\
\end{tabular}
\end{ruledtabular}\end{table}\endgroup

\begin{figure}[ptb]
\includegraphics[width=0.24\textwidth, clip]{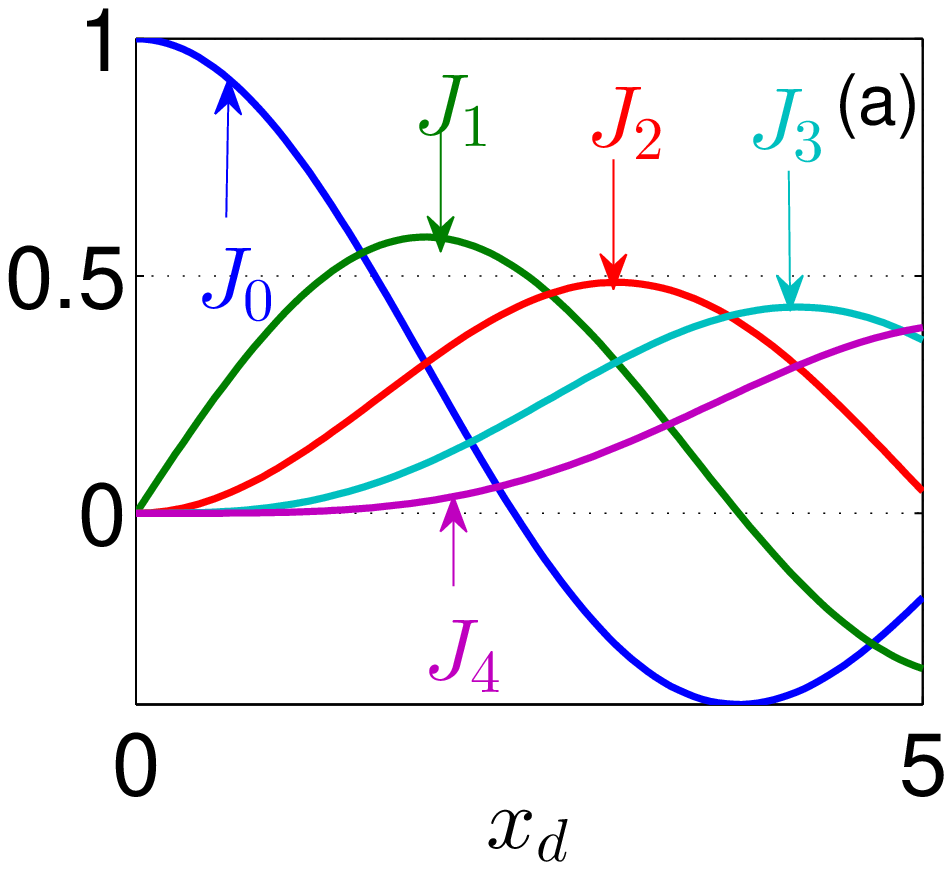}\includegraphics[width=0.24\textwidth, clip]{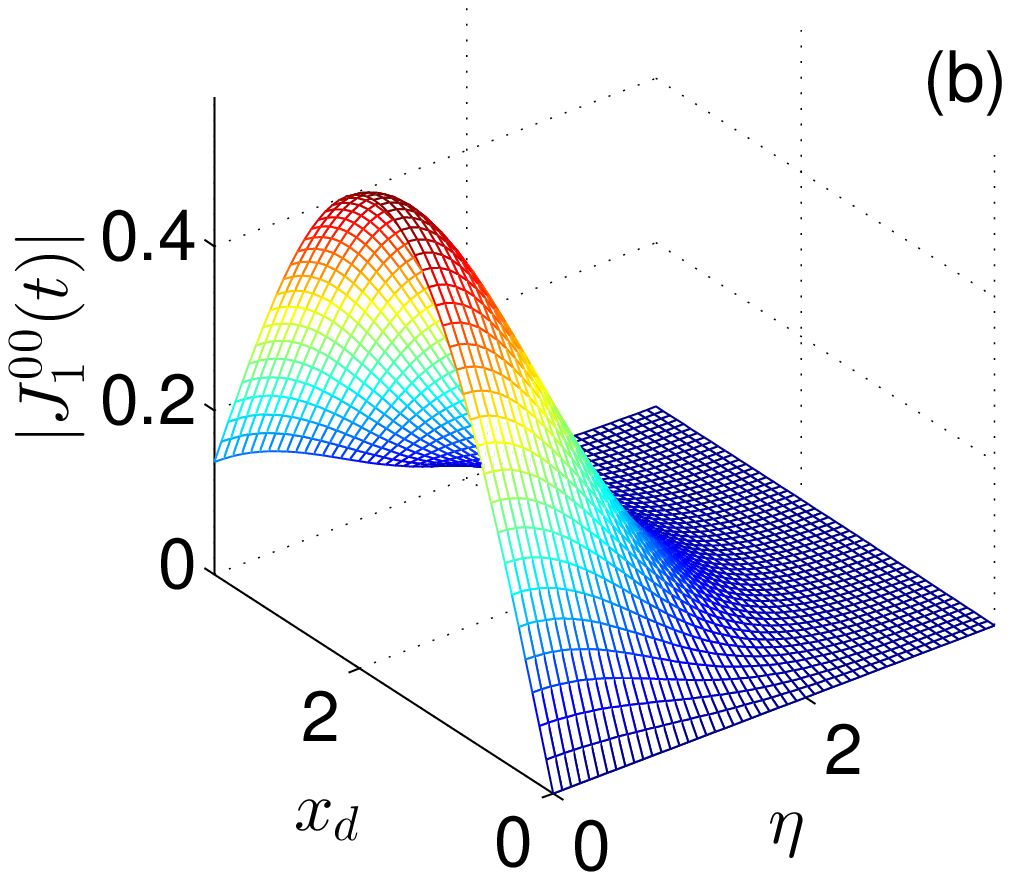}\newline%
\includegraphics[width=0.24\textwidth, clip]{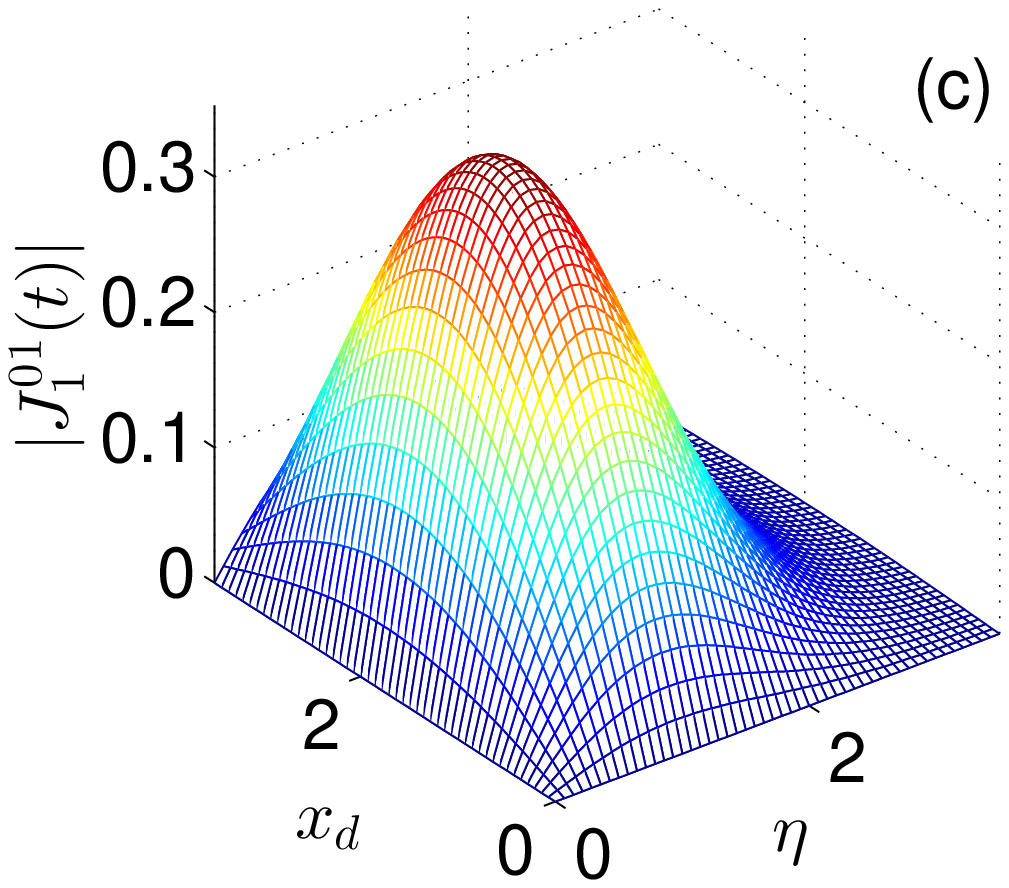}\includegraphics[width=0.24\textwidth, clip]{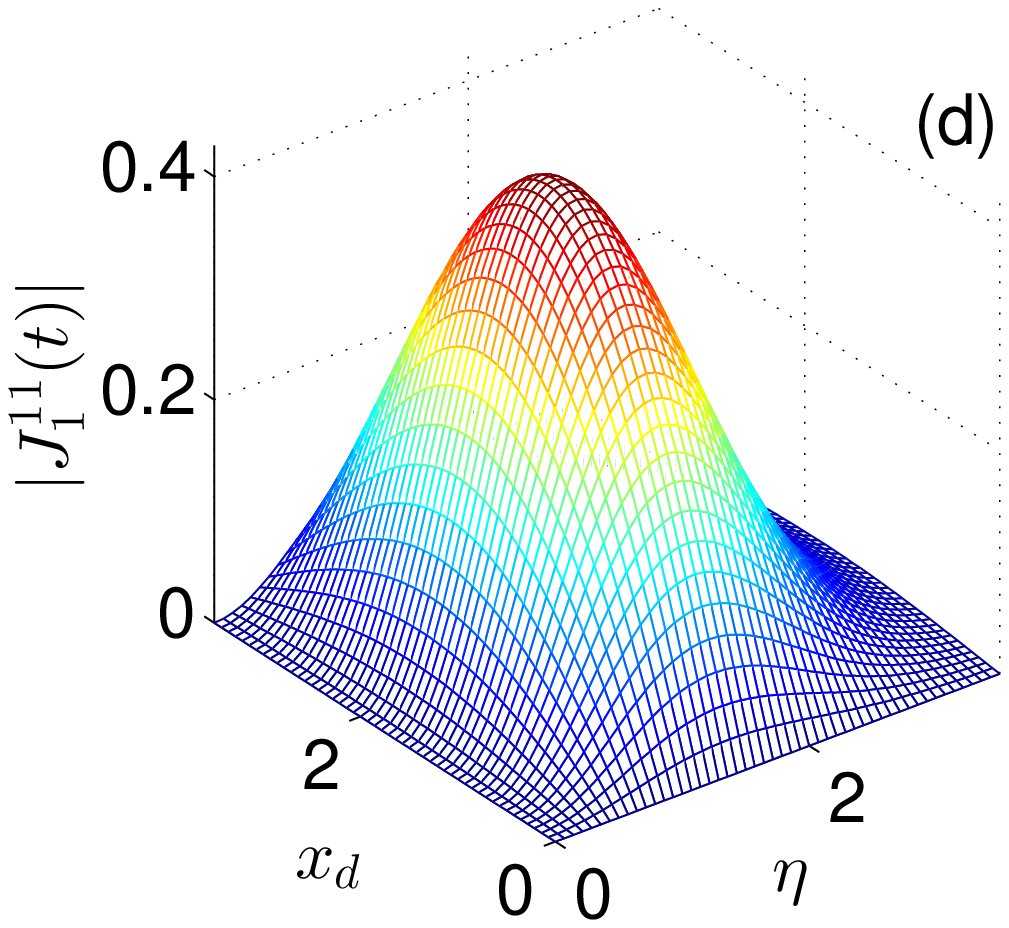}
\caption{(Color online) Bessel functions $J_{N}(x_{d})$ of the first
kind\textbf{,} with $N=0,\;1,\;2,\;3$\textbf{, }are plotted as\textbf{
}functions of the ratio $x_{d}$ in (a). $\left\vert J_{1}^{00}(t)\right\vert
$, $\left\vert J_{1}^{01}(t)\right\vert $, and $\left\vert J_{1}%
^{11}(t)\right\vert $ have been plotted in (b), (c), and (d) as functions of
$\eta$ and $x_{d}$, respectively. Recall that $x_{d}=2\Omega_{d}/\omega_{d}$
is the ratio between the driving field-SQ Rabi frequency $\Omega_{d}$ and the
frequency $\omega_{d}$ of the driving field.}%
\label{fig2}%
\end{figure}

\subsection{Bessel functions and coupling strengths}

In the process of generating nonclassical photon states, the coupling strength $J_{N}^{mn}(t)$
plays an important role. In our study here, the Bessel functions of the first
kind are crucial factors in the coupling strengths. The possible values of the
Bessel functions depend on the ratio $x_{d}$ between the driving field-SQ Rabi
frequency $\Omega_{d}$ and the frequency $\omega_{d}$ of the driving field.
For several recent experiments with superconducting quantum circuits, the
coupling constant $\Omega_{d}$ is usually in the range from several tens of
MHz to several hundreds of MHz, e.g., $10\,\mathrm{MHz}<\Omega_{d}<500$ MHz.
The frequency $\omega_{d}$ of the driving field is in the range of GHz, e.g.,
$1\,\mathrm{GHz}\leq\omega_{d}\leq20$ GHz. Thus the ratio $x_{d}$ is in the
range
\begin{equation}
10^{-9}\leq x_{d}\leq1.
\end{equation}
For completeness and to allow a comparison between them, several Bessel
functions are plotted as a function of the parameter $x_{d}=2\Omega_{d}%
/\omega_{d}$ in Fig.~\ref{fig2}(a), which clearly shows $J_{0}(x_{d}%
)>J_{1}(x_{d})>J_{2}(x_{d})>\cdots>J_{N}(x_{d})$ in the range of $10^{-9}\leq
x_{d}\leq1$. Thus if the classical driving field is chosen such that the ratio
$x_{d}$ is less than $0.5$, then we only need to consider the terms in the
Hamiltonian in Eq.~(\ref{eq:3}) with the Bessel functions $J_{0}(x_{d})$ and
$|J_{1}(x_{d})|=|J_{-1}(x_{d})|$, and other terms are negligibly small. As
discussed above, it should be noted that the frequency $\omega_{d}$ of the
driving field has no effect on the coupling between the SQ and the quantized
field in terms of the Bessel function $J_{0}(x_{d})$. Thus the
driving-field-assisted transitions between the SQ and the quantized field are
determined by the terms with the Bessel functions $J_{\pm1}(x_{d})$,
when other high-order Bessel functions are neglected.
Figure~\ref{fig2}(a) also shows that the terms with the
Bessel function $J_{\pm2}(x_{d})$ are also not negligible when $x_{d}$ becomes
larger, e.g., $0.5\leq x_{d}<1$. Thus in the regime $0\leq x_{d}<1$, the
terms with high-order Bessel functions (e.g., the ones with $N\geq3$) can be neglected.

As an example, Figs.~\ref{fig2}(b, c, d) illustrate how $\left\vert J_{1}%
^{mn}(t)\right\vert $ are affected by $m$, $n$, $x_{d}$, and $\eta$. Since the
maximal point occurs at $\eta=\sqrt{m+n}$\textbf{, }if other variables are
fixed, thus we can find an obvious shift of the maximal point along the $\eta
$-axis with increasing $m+n$. We can also find that $\left\vert J_{N}%
^{mn}(t)\right\vert $ have similar results as those of $\left\vert J_{1}%
^{mn}(t)\right\vert $ versus $m$, $n$, $x_{d}$, and $\eta$. By tuning $x_{d}$
and $\eta$, we can change the Rabi frequencies, and thus optimize the
generation time.

\section{Generating non-classical photon states using superconducting quantum
circuits}

In this section, we discuss how to generate non-classical photon states
via transverse and longitudinal couplings between SQs and the single-mode
cavity field\textbf{,} with the assistance of a classical driving field.

\subsection{Interaction Hamiltonian and time-evolution operators}

Let us now analyze the interaction Hamiltonian and time evolutions for the
three different processes based on the Hamiltonian in Eq.~(\ref{eq:3}). We
have three different interaction Hamiltonians. In the interaction picture with
the resonant conditions of different photon processes, by assuming $n=m+k$ for
the red-sideband excitation, $m=n+k$ for the blue-sideband excitation, and
$m=n$ for the carrier process. We now discuss the general case for coupling
constants with any number of Bessel functions. For a red process with the
$N$th Bessel functions, we derive the Hamiltonian
\begin{equation}
H_{r}=J_{N,r}^{\left(  k\right)  }\sum_{m}\frac{(-1)^{m}\eta^{2m}\sigma
_{+}a^{\dagger m}a^{m+k}}{m!(m+k)!}+\mathrm{h.c.}, \label{eq:7}%
\end{equation}
with the resonant condition
\[
N\omega_{d}=\omega_{z}-k\omega.
\]
For a blue process with the $N$th Bessel functions, we have the Hamiltonian
\begin{equation}
H_{b}=J_{N,b}^{\left(  k\right)  }\sum_{n}\frac{(-1)^{n}\eta^{2n}\sigma
_{+}a^{\dagger(n+k)}a^{n}}{n!(n+k)!}+\mathrm{h.c.}, \label{eq:8}%
\end{equation}
with the resonant condition
\[N\omega_{d}=\omega_{z}+k\omega.\]
The parameters $J_{N,r}^{k}$ and $J_{N,b}^{k}$ for the red process in
Eq.~(\ref{eq:7}) and the blue one in Eq.~(\ref{eq:8}) are given by
\begin{align}
\label{eq:20-1} J_{N,r}^{\left(  k\right)  }  &  =(-1)^{k}\frac{\omega_{x}}{2}J_{N}\exp\left[
-\frac{1}{2}\left(  \frac{2g}{\omega}\right)  ^{2}+iN\phi_{d}^{\left(
\beta\right)  }\right]  \eta^{k},\\
\label{eq:21-1} J_{N,b}^{\left(  k\right)  }  &  =\frac{\omega_{x}}{2}J_{N}\exp\left[
-\frac{1}{2}\left(  \frac{2g}{\omega}\right)  ^{2}+iN\phi_{d}^{\left(
\beta\right)  }\right]  \eta^{k},
\end{align}
where the subscript $\beta$ takes either $r$ or $b$, we use $\phi_{0}^{\left(
\beta\right)  }$ to characterize the initial phase of either the red or the
blue process. For the carrier process with the $N$th Bessel functions, the
interaction Hamiltonian is given by%

\begin{equation}
H_{c}=J_{N,c}^{\left(  0\right)  }\sum_{n}\eta^{2n}\frac{(-1)^{n}\sigma
_{+}a^{\dagger n}a^{n}}{n!n!}+\mathrm{h.c.}, \label{eq:9}%
\end{equation}
with the resonant condition
\[
N\omega_{d}=\omega_{z},
\]
and the coupling constant
\begin{equation}
J_{N,c}^{\left(  0\right)  }=\frac{1}{2}\omega_{x}J_{N}\exp\left[  -\frac
{1}{2}\eta^{2}+iN\phi_{d}^{\left(  c\right)  }\right]  . \label{eq:23-1}
\end{equation}
We also note that all non-resonant terms have been neglected when
Eqs.~(\ref{eq:7}-\ref{eq:9}) are derived. The dynamical evolutions of the
systems corresponding to these three different processes can be described via
time-evolution operators. For example, for the $k$th red, blue, and carrier
sideband excitations, we respectively have the evolution operators%
\begin{align}
U_{N,r}^{\left(  k\right)  }\left(  t\right)  =  &  \sum_{n=0}^{k-1}\left\vert
n\right\rangle \left\langle n\right\vert \sigma_{00}+\sum_{n=0}^{\infty}%
\cos\left(  \left\vert \Omega_{N,r}^{k,n}\right\vert t\right)  \left\vert
n\right\rangle \left\langle n\right\vert \sigma_{11}\nonumber\\
&  +\sum_{n=0}^{\infty}e^{-i\phi_{N,r}^{k,n}-i\pi/2}\mathrm{sin}\left(
\left\vert \Omega_{N,r}^{k,n}\right\vert t\right)  \left\vert n+k\right\rangle
\left\langle n\right\vert \sigma_{-}\nonumber\\
&  +\sum_{n=0}^{\infty}e^{i\phi_{N,r}^{k,n}-i\pi/2}\mathrm{sin}\left(
\left\vert \Omega_{N,r}^{k,n}\right\vert t\right)  \left\vert n\right\rangle
\left\langle n+k\right\vert \sigma_{+}\nonumber\\
&  +\sum_{n=0}^{\infty}\cos\left(  \left\vert \Omega_{N,r}^{k,n}\right\vert
t\right)  \left\vert n+k\right\rangle \left\langle n+k\right\vert \sigma_{00},
\label{eq:Ur}%
\end{align}%
\begin{align}
U_{N,b}^{\left(  k\right)  }\left(  t\right)  =  &  \sum_{n=0}^{k-1}\left\vert
n\right\rangle \left\langle n\right\vert \sigma_{11}+\sum_{n=0}^{\infty}%
\cos\left(  \left\vert \Omega_{N,b}^{k,n}\right\vert t\right)  \left\vert
n\right\rangle \left\langle n\right\vert \sigma_{00}\nonumber\\
&  +\sum_{n=0}^{\infty}e^{i\phi_{N,b}^{k,n}-i\pi/2}\mathrm{sin}\left(
\left\vert \Omega_{N,b}^{k,n}\right\vert t\right)  \left\vert n+k\right\rangle
\left\langle n\right\vert \sigma_{+}\nonumber\\
&  +\sum_{n=0}^{\infty}e^{-i\phi_{N,b}^{k,n}-i\pi/2}\mathrm{sin}\left(
\left\vert \Omega_{N,b}^{k,n}\right\vert t\right)  \left\vert n\right\rangle
\left\langle n+k\right\vert \sigma_{-}\nonumber\\
&  +\sum_{n=0}^{\infty}\cos\left(  \left\vert \Omega_{N,b}^{k,n}\right\vert
t\right)  \left\vert n+k\right\rangle \left\langle n+k\right\vert \sigma_{11},
\label{eq:Ub}%
\end{align}
and%
\begin{align}
U_{N,c}^{\left(  0\right)  }\left(  t\right)  =  &  \sum_{n=0}^{\infty}%
\cos\left(  \left\vert \Omega_{N,c}^{0,n}\right\vert t\right)  \left\vert
n\right\rangle \left\langle n\right\vert \sigma_{11}\nonumber\\
&  +\sum_{n=0}^{\infty}e^{-i\phi_{N,c}^{0,n}-i\pi/2}\mathrm{sin}\left(
\left\vert \Omega_{N,c}^{0,n}\right\vert t\right)  \left\vert n\right\rangle
\left\langle n\right\vert \sigma_{-}\nonumber\\
&  +\sum_{n=0}^{\infty}e^{i\phi_{N,c}^{0,n}-i\pi/2}\mathrm{sin}\left(
\left\vert \Omega_{N,c}^{0,n}\right\vert t\right)  \left\vert n\right\rangle
\left\langle n\right\vert \sigma_{+}\nonumber\\
&  +\sum_{n=0}^{\infty}\cos\left(  \left\vert \Omega_{N,c}^{0,n}\right\vert
t\right)  \left\vert n\right\rangle \left\langle n\right\vert \sigma_{00},
\label{eq:Uc}%
\end{align}
where the complex Rabi frequency and its phase angle are respectively defined
as%
\begin{align}
\Omega_{N,\beta}^{k,n}  &  =J_{N,\beta}^{(k)}\sqrt{\frac{n!}{\left(  n+k\right)
!}}L_{n}^{\left(  k\right)  }\left(  \eta^{2}\right)  ,\label{eq:Omegab}\\
\phi_{N,\beta}^{k,n}  &  =\arg\left(  \Omega_{N,\beta}^{k,n}\right) ,
\end{align}
with $\beta=r, b, c$ and $J_{N,\beta}^{(k)}$ are given in Eqs.~(\ref{eq:20-1}), (\ref{eq:21-1}), and (\ref{eq:23-1}).  Here $L_{n}^{\left(  k\right)  }\left(  x\right)  $ represents the generalized
Laguere polynomials. Let us assume that the two eigentates $|g\rangle$ and
$|e\rangle$ of the Pauli operator $\sigma_{z}$ satisfy $\sigma_{z}\left\vert
g\right\rangle =-\left\vert g\right\rangle $, and $\sigma_{z}\left\vert
e\right\rangle =\left\vert e\right\rangle $, then we define the following
operators $\sigma_{ii}$ as $\sigma_{00}=\left\vert g\right\rangle \left\langle
g\right\vert $, $\sigma_{11}=\left\vert e\right\rangle \left\langle
e\right\vert $, $\sigma_{01}=\left\vert g\right\rangle \left\langle
e\right\vert $, and $\sigma_{10}=\left\vert e\right\rangle \left\langle
g\right\vert $.

\subsection{Synthesizing nonclassical photon states}

We find that the interaction Hamiltonians in Eqs.~(\ref{eq:7}-\ref{eq:9}) in
the displacement picture are very similar to those for trapped ions~\cite{wei}%
. Therefore, in principle the non-classical photon states can be generated by
alternatively using the above three different controllable processes. We
expect that the prepared target state is
\begin{equation}
\left\vert \psi_{n_{\max}}\right\rangle =\sum_{n=0}^{n_{\max}}C_{n}\left\vert
n\right\rangle \otimes\left\vert q\right\rangle , \label{eq:27}%
\end{equation}
where $n_{\max}$ is a maximal photon number in the photon state of the target
state. Here\textbf{, }$\left\vert n\right\rangle \otimes\left\vert
q\right\rangle \equiv\left\vert n\right\rangle \left\vert q\right\rangle $
denotes that the cavity field is in the photon number state $|n\rangle$ and
the qubit is in the state $|q\rangle$, which can be either the
ground $|g\rangle$ or excited $|e\rangle$ state. The
parameter $|C_{n}|^{2}$ is the probability of the state $\left\vert
n\right\rangle \otimes\left\vert q\right\rangle $. The steps for producing the
target state for both the case $q=g$ and $q=e$ are very similar. We thus take
$q=g$ as an example to present the detailed steps. The target state then takes
the form
\begin{equation}
\left\vert \psi_{n_{\max}}\right\rangle =\sum_{k=0}^{n_{\max}}C_{k}\left\vert
k\right\rangle \otimes\left\vert g\right\rangle . \label{eq:psif}%
\end{equation}
We point out that all the states here (e.g., the target state) are observed in
the displacement picture\textbf{, }if we do not specify this.

We assume that the system is initially in the state $\left\vert 0\right\rangle
\left\vert g\right\rangle $. Then, by taking similar steps as in
Ref.~\cite{wei}, we can generate an arbitrary state in which the states
$|\psi_{n}\rangle$ in the $n$th step and $|\psi_{n-1}\rangle$ in the $(n-1)$th
step have the following relation,
\begin{equation}
\left\vert \psi_{n}\right\rangle =U_{0}^{(n)\dag}\left(  t_{n}\right)
U_{N,\beta_{n}}^{(n)}\left(  t_{n}\right)  U_{0}^{(n)}\left(  0\right)
\left\vert \psi_{n-1}\right\rangle , \label{eq:psi_recur}%
\end{equation}
with
\begin{equation}
\left\vert \psi_{n}\right\rangle =\sum_{k=0}^{n}C_{kg}^{\left(  n\right)
}\left\vert k\right\rangle \left\vert g\right\rangle +C_{0e}^{\left(
n\right)  }\left\vert 0\right\rangle \left\vert e\right\rangle .
\label{eq:psi_n_expand}%
\end{equation}
Here if $n=n_{\max}$, $\ C_{0e}=0$, and $C_{kg}^{\left(  n\right)  }=C_{k}$,
then $\left\vert \psi_{n}\right\rangle $ in Eq.~(\ref{eq:psi_n_expand}) is
reduced to $\left\vert \psi_{n_{\max}}\right\rangle $ in Eq.~(\ref{eq:psif}).
Above, $t_{n}$ is the time duration of the control pulse for the $n$th step.
The unitary transform $U_{0}^{(n)}\left(  t\right)  $ is defined as
\begin{equation}
U_{0}^{(n)}\left(  t\right)  =V_{0}\left(  t\right)  U_{d}^{\left(  n\right)
}\left(  t\right)  . \label{eq:U0}%
\end{equation}
Here $V_{0}\left(  t\right)  $ is given in Eq. (\ref{eq:V0}). Also,
$U_{d}^{\left(  n\right)  }\left(  t\right)  $ is actually $U_{d}\left(
t\right)  $ in Eq. (\ref{eq:Ud}), but with $\omega_{d}$ and $\phi_{d}$
replaced by $\omega_{d}^{\left(  n\right)  }$ and $\phi_{d}^{\left(  n\right)
}$, which denote respectively the frequency and phase of the driving field for
the $n$th step. Moreover, $U_{N,\beta_{n}}^{(n)}\left(  t_{n}\right)  $
denotes a unitary transform of the $n$th step, and is taken from one of
Eq.~(\ref{eq:Ur}), Eq.~(\ref{eq:Ub}), and Eq.~(\ref{eq:Uc}) depending on which
one is chosen as $\beta_{n}$ among the characters \textquotedblleft$r$",
\textquotedblleft$b$", and \textquotedblleft$c$". In $U_{N,\beta_{n}}%
^{(n)}\left(  t_{n}\right)  $, the parameters $\omega_{d}$ and $\phi_{d}$ must
also be replaced by $\omega_{d}^{\left(  n\right)  }$ and $\phi_{d}^{\left(
n\right)  }$, respectively.

The target of the $n$th step is to generate the state $\left\vert \psi
_{n}\right\rangle $ from the state $\left\vert \psi_{n-1}\right\rangle $. We
assume $\left\vert \psi_{n-1}\right\rangle $ is in the displacement picture,
which is the state generated after the $(n-1)$th step. We first use
$U_{0}^{(n)}\left(  t_{n}\right)  $ to transfer $\left\vert \psi
_{n-1}\right\rangle $ from the displaced picture into the interaction picture.
Then we choose one of the evolution operators in Eqs.~(\ref{eq:Ur})--(\ref{eq:Uc}) with a proper photon number to reach
the target state in Eq.~(\ref{eq:psi_recur}). Since the state $\left\vert
\psi_{n}\right\rangle $ should also be represented in the displaced picture,
after the state of the $n$th step via the evolution operators $U_{N,\beta_{n}%
}^{(n)}\left(  t_{n}\right)  $ and $U_{0}^{(n)}\left(  0\right)  $, we have to
transfer it back to the displaced picture, which results in the appearance of
$U_{0}^{(n)\dag}\left(  t_{n}\right)  $ in Eq.~(\ref{eq:psi_recur}).

\begin{figure}[ptb]
\includegraphics[ width=0.48\textwidth, clip]{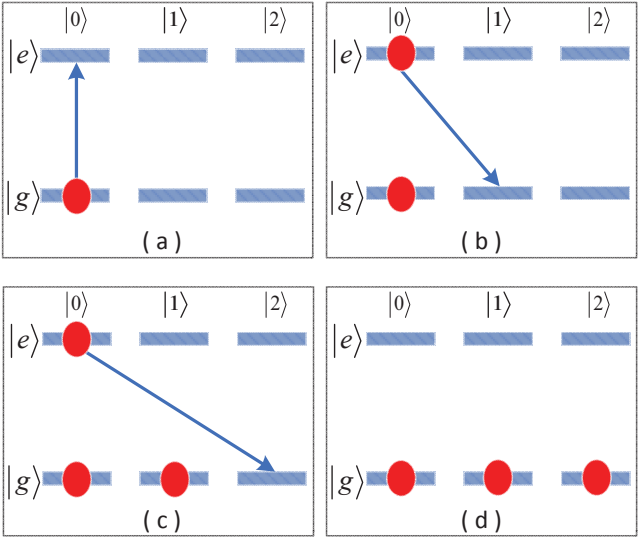}\caption{(Color online)
Schematic diagram for state generation of the target state $\left\vert
\psi_{n_{\max}}\right\rangle =\sum_{k=0}^{n_{\max}}C_{k}\left\vert
k\right\rangle \otimes\left\vert g\right\rangle $, for example, $n_{\max}=2$.
(a) Step 0, the system is initially in the state $\left\vert 0\right\rangle
\otimes\left\vert g\right\rangle $, and the arrow means that a carrier process is
applied with the $0$-photon inside the cavity, denoted by the operator
$U_{N,c}^{(0)}\left(  t_{0}\right)  $. (b) Step 1, after the step $0$, the
system is in the state $C_{0g}^{\left(  0\right)  }\left\vert 0\right\rangle
\otimes\left\vert g\right\rangle +C_{0e}^{\left(  0\right)  }\left\vert
0\right\rangle \otimes\left\vert e\right\rangle$, in which the parameters
$C_{0e}^{\left(  0\right)  }$ and $C_{0g}^{\left(  0\right)  }$ are determined
by the time duration $t_{0}$. The arrow means that a $1$-photon red
process is applied with the time evolution operator $U_{N,r}^{\left(
1\right)  }\left(  t_{1}\right)  $ after the step $0$. (c) Step 2, after the
step $1$, the system is in the state $C_{0g}^{\left(  1\right)  }\left\vert
0\right\rangle \otimes\left\vert g\right\rangle +C_{1g}^{\left(  1\right)
}\left\vert 1\right\rangle \otimes\left\vert g\right\rangle +C_{0e}^{\left(
1\right)  }\left\vert 0\right\rangle \otimes\left\vert e\right\rangle
$, in which the coefficients of the superposition are determined by
the time duration $t_{0}$ and $t_{1}$. The arrow means that a $2$-photon red
process $U_{N,r}^{\left(  2\right)  }\left(  t_{2}\right)  $ is applied to the
system after the step $1$. (d) After the step $2$ \textbf{(}with well-chosen
time durations $t_{0}$, $t_{1}$ and $t_{2}$\textbf{)}, the system is in the
state $C_{0}\left\vert 0\right\rangle \otimes\left\vert g\right\rangle
+C_{1}\left\vert 1\right\rangle \otimes\left\vert g\right\rangle
+C_{2}\left\vert 2\right\rangle \otimes\left\vert g\right\rangle $, which is
just the target state $\left\vert \psi_{2}\right\rangle $. Other
superpositions can also be generated using similar steps.}%
\label{fig3}%
\end{figure}

The \textit{longitudinal} coupling results in multi-photon processes.
Thus the state preparation using the longitudinal coupling is in principle
more convenient than that using a single-photon transition in the usual
Jaynes-Cumming model~\cite{liuepl}. For example, the Fock state $|n\rangle$
can be generated with a carrier process and a longitudinal coupling
field-induced $n$-photon process. However, it needs $2n$ steps ($n$ step
carrier and $n$ step red-sideband processes) to produce a Fock state
$|n\rangle$ if we use the Jaynes-Cumming model~\cite{liuepl}. The selection of
$U_{N,\beta_{n}}^{\left(  n\right)  }\left(  t_{n}\right)  $ in
Eq.~(\ref{eq:psi_recur}) for each step is almost the same as that in
Ref.~\cite{wei}. That is, the target state in Eq.~(\ref{eq:psif}) can be
obtained either by virtue of one carrier process and $n_{\mathrm{max}}$ red-sideband excitations, or by virtue of
one carrier process and $n_{\mathrm{max}}$ blue-sideband excitations.

The steps to generate the target state in Eq.~(\ref{eq:psif}) from the initial
state $|0\rangle|g\rangle$ using carrier and red sideband excitations are
schematically shown in Fig.~\ref{fig3} using a simple example. All steps for
the required unitary transformations are described as follows. First\textbf{,}
the initial state $\left\vert 0\right\rangle \left\vert g\right\rangle $ is
partially excited to $\left\vert 0\right\rangle \left\vert e\right\rangle $ by
a carrier process ($n=0$) with a time duration $t_{0}$ such that the
probability $|C_{0g}^{(0)}|^{2}$ in $\left\vert 0\right\rangle \left\vert
g\right\rangle $ satisfies the condition $|C_{0g}^{(0)}|^{2}=|C_{0}|^{2}$, with
$C_{0}$ given in Eq.~(\ref{eq:psif}). After the carrier process, the driving
fields are sequentially applied to the qubit with $n_{\mathrm{max}}$ different
frequency matching conditions, such that a single-photon, two-photon, until
$n_{\mathrm{max}}$-photon red processes can occur. Thus the subscript
$\beta_{n}$ in the unitary transform $U_{N,\beta_{n}}^{(n)}$ satisfies the
conditions $\beta_{n}=c$ with $n=0$\textbf{, }and $\beta_{n}=r$ for $n\geq1$.
By choosing appropriate time durations and the phases of the driving fields in
each step, which in principle can be obtained using Eq.~(\ref{eq:psi_recur}),
we can obtain the target state shown in Eq.~(\ref{eq:psif}). The detailed
descriptions can be found in Appendix~\ref{Append:Opt1}.

\section{The initial state and target state}

\begin{figure}[ptb]
\includegraphics[width=0.42\textwidth, clip]{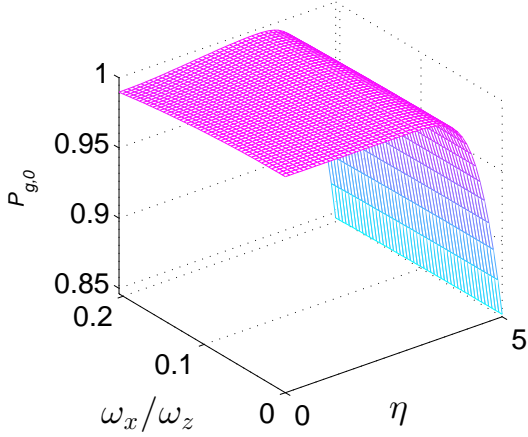} \caption{(Color online)
Probability $P_{g,0}$ for the ground state to be the vacuum state in the
displacement picture as a function of $\eta$ and $\omega_{x}/\omega_{z}$. Here
we assume $\omega_{z}/2\pi=19.5\operatorname{GHz}$, and $\omega
=2\operatorname{GHz}$. Recall that $\eta$ is the normalized coupling or the
Lamb-Dicke parameter.}%
\label{fig4}%
\end{figure}

Above, we assumed that the target state is generated from the initial state
which is the vacuum state in the displacement picture defined by
Eq.~(\ref{eq:V}). However, in experiments, the initial state is usually the
ground state\textbf{, }obtained by cooling the sample inside a
dilution refrigerator. We now investigate the ground state of the effective
Hamiltonian when there is no driving field. The Hamiltonian $H_{\mathrm{eff}}$
without driving field can be expressed as%
\begin{equation}
H_{\mathrm{eff}}^{\prime}=\frac{\hbar}{2}\omega_{z}\sigma_{z}+\hbar\omega
a^{\dagger}a+\frac{\hbar}{2}\omega_{x}\left\{  \sigma_{+}\exp\left[
\eta\left(  a^{\dagger}-a\right)  \right]  +\mathrm{h.c.}\right\}  ,
\label{eq:Heff_p}%
\end{equation}
in the displacement picture. However, in the original picture, the
corresponding Hamiltonian is%
\begin{equation}
H^{\prime}=\frac{\hbar}{2}\omega_{z}\sigma_{z}+\frac{\hbar}{2}\omega_{x}%
\sigma_{x}+\hbar\omega a^{\dagger}a+\hbar g\sigma_{z}(a+a^{\dagger}),
\label{eq:34}%
\end{equation}
which possesses the characteristics of broken-symmetry and strong coupling and
is hence difficult to solve analytically. Due to the mathematical equivalence
between Eq.~(\ref{eq:Heff_p}) and Eq.~(\ref{eq:34}), it is also difficult to
solve Eq.~(\ref{eq:Heff_p}) analytically. We thus resort to numerical
calculations to obtain the ground state of $H_{\mathrm{eff}}^{\prime}$.
We define the ground state of the Hamiltonian $H_{\mathrm{eff}%
}^{\prime}$ as $\left\vert \psi_{g}\right\rangle $, and the probability of the
ground state $\left\vert \psi_{g}\right\rangle $ to be in the vacuum state
$\left\vert 0\right\rangle $ as $P_{g,0}$. The relation between $\left\vert
\psi_{g}\right\rangle $ and $P_{g,0}$ can be written as
\begin{align}
\left\vert \psi_{g}\right\rangle  &  =\left[  \xi_{0}\left\vert 0\right\rangle
+\sqrt{1-|\xi_{0}|^{2}}\left\vert \delta\psi_{g}\right\rangle \right]
|g\rangle,\label{eq:psi_g}\\
P_{g,0}  &  =\left\vert \xi_{0}\right\vert ^{2},
\end{align}
where $|\delta\psi_{g}\rangle$ denotes a superposition of photon number states
except the vacuum state. In Fig.~\ref{fig4}, as an example, by taking
$\omega_{z}/2\pi=19.5$ GHz and $\omega=2$ GHz, we have plotted $P_{g,0}$ as a
function of $\eta$ and $\omega_{x}$. We find that the probability $P_{g,0}%
\geq0.99$\textbf{, }at least in the region $0<\eta<3.5$ and $0<\omega
_{x}/\omega_{z}<0.2$. More specifically, the ground state of the Hamiltonian in
Eq.~(\ref{eq:Heff_p}) is closer to the vacuum state when the parameters $\eta$
and $\omega_{x}$ are smaller. Thus, our assumption that the initial state of
the cavity field in the displacement picture is the vacuum state, can always
be valid only if the related parameters, such as $\omega_{x}$ and $\eta$, are
properly chosen.

We have demonstrated how to generate an arbitrary superposition of different
Fock sates from the vacuum state in the displacement picture. Thus, once the
state is generated, we have to displace the generated state back to the
original picture via the displacement operator $D^{\dagger}\left(  \eta
\sigma_{z}/2\right)  $. For example, the initial state $|0\rangle|g\rangle$ in
the displacement picture becomes
\begin{equation}
D^{\dag}\left(  \frac{\eta}{2}\sigma_{z}\right)  |0\rangle|g\rangle=\left\vert
\frac{\eta}{2},0\right\rangle \left\vert g\right\rangle ,
\end{equation}
in the original picture, where $\left\vert \alpha,n\right\rangle =$ $D\left(
\alpha\right)  \left\vert n\right\rangle $ denotes the displaced number
state~\cite{Oliveira1990}. Similarly, the target state $\left\vert
\psi_{n_{\max}}\right\rangle $ in the displacement picture becomes
\begin{equation}
\left\vert \psi_{n_{\max}}^{D}\right\rangle =\sum_{n=0}^{n_{\max}}%
C_{n}\left\vert \frac{\eta}{2},n\right\rangle \left\vert g\right\rangle ,
\label{eq:33}%
\end{equation}
in the original picture. It is obvious that the initial state $\left\vert
\eta/2,0\right\rangle $ of the cavity field in the original picture is a
coherent state with the average photon number $\left(  \eta/2\right)  ^{2}$,
while the target state is the superposition of the displaced number states.

\begin{figure}[ptb]
\includegraphics[width=0.48\textwidth,clip]{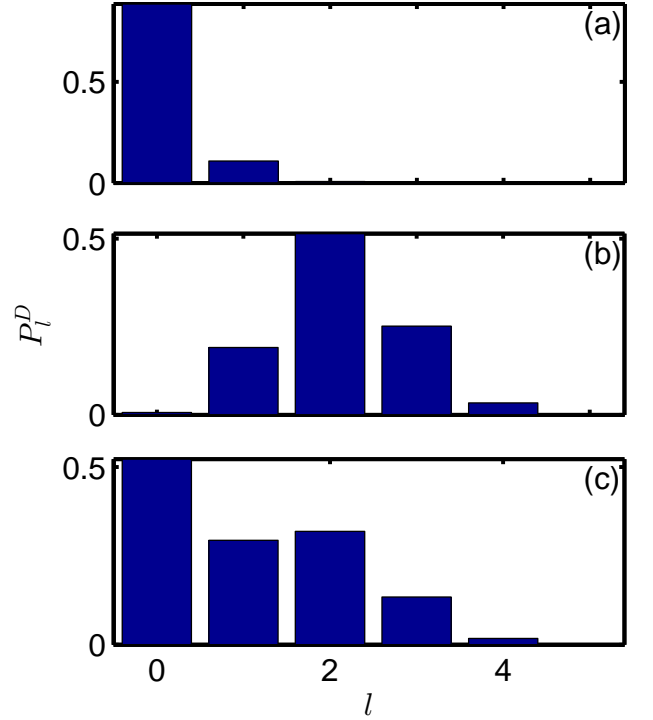} \newline\caption{Photon
number distributions of (a) $\left\vert \eta/2,0\right\rangle $, (b)
$\left\vert \eta/2,2\right\rangle $, and (c) $\left(  \left\vert
\eta/2,0\right\rangle +\left\vert \eta/2,2\right\rangle \right)  /\sqrt{2}$.
Here $l$ refers to the photon number and $P^D_l$ refers to the probability on $\left\vert l\right\rangle$. We have taken the Lamb-Dicke parameter $\eta=2g/\Omega=0.7.$}%
\label{fig5}%
\end{figure}

The statistical properties of a displaced number state with $D(\alpha
)|n\rangle$ can be described by the probabilities of the photon number
distribution as below
\begin{equation}
\left\langle l|\alpha,n\right\rangle =\left\{
\begin{array}
[c]{ll}%
\frac{\alpha^{l-n}\sqrt{n!/l!}}{\exp\left(  \left\vert \alpha\right\vert
^{2}/2\right)  }L_{n}^{\left(  l-n\right)  }\left(  \left\vert \alpha
\right\vert ^{2}\right)  , & \,l\geq n,\\
\\
\frac{\left(  -\alpha^{\ast}\right)  ^{\left(  n-l\right)  }\sqrt{l!/n!}}%
{\exp\left(  \left\vert \alpha\right\vert ^{2}/2\right)  }L_{l}^{\left(
n-l\right)  }\left(  \left\vert \alpha\right\vert ^{2}\right)  , & \,l<n.
\end{array}
\right.
\end{equation}
Thus the displaced target state in Eq.~(\ref{eq:33}) can be written as%
\begin{align}
\left\vert \psi_{n_{\max}}^{D}\right\rangle  &  =\sum_{n=0}^{n_{\max}}%
C_{n}\sum_{l=0}^{\infty}\left\vert l\right\rangle \left\langle l\right.
\left\vert \frac{\eta}{2},n\right\rangle \left\vert g\right\rangle \nonumber\\
&  =\sum_{l=0}^{\infty}C_{l}^{D}\left\vert l\right\rangle \left\vert
g\right\rangle ,
\end{align}
where $C_{l}^{D}=\sum_{n=0}^{n_{\max}}C_{n}\left\langle l|\eta
/2,n\right\rangle $. The probability of the target sate $\left\vert
\psi_{n_{\max}}^{D}\right\rangle $ to be in the photon number $\left\vert
l\right\rangle $ in the original picture can be given as
\begin{equation}
P_{l}^{D}=\left\vert C_{l}^{D}\right\vert ^{2}=\sum_{n=0}^{n_{\max}}\sum
_{m=0}^{n_{\max}}C_{n}C_{m}^{\ast}\Big\langle l  \left\vert \frac
{\eta}{2},n\right\rangle \left\langle \frac{\eta}{2},m\right\vert
l\Big\rangle . \label{eq:P_D_l}%
\end{equation}
In Fig.~\ref{fig5}, as an example, the distribution probabilities $P_{l}^{D}$
are plotted for different photon states, that is, $\left\vert \psi_{n_{\max}%
}^{D}\right\rangle $ is taken as $\left\vert \eta/2,0\right\rangle \left\vert
g\right\rangle ,\left\vert \eta/2,2\right\rangle \left\vert g\right\rangle ,$
or $\left(  \left\vert \eta/2,0\right\rangle +\left\vert \eta/2,2\right\rangle
\right)  \left\vert g\right\rangle /\sqrt{2}$, which is $|0\rangle$,
$|2\rangle$ and $(|0\rangle+|2\rangle)/\sqrt{2}$, respectively, in the
displacement picture. Figure~\ref{fig5} shows that the photon number states in
the displacement picture are redistributed after these states are sent back to
the original picture. Even though the state in Fig.~\ref{fig5}(c) is the linear sum
of the states in Fig.~\ref{fig5}(a) and Fig.~\ref{fig5}(b), the photon
number distributions are not linearly additive. Because the interference
between different displaced number states, which corresponds to the terms of
$m\neq n$ in Eq.~(\ref{eq:P_D_l}), can also give rise to the variation of the
photon number distribution. It is clear that a number state in the displaced
picture can become a superposition of number states in the original picture,
which might offer a convenient way to prepare nonclassical photon states.

\section{Numerical analysis}

We have presented a detailed analysis on how to prepare nonclassical photon
states using the longitudinal-coupling-induced multi-photon processes in an
ideal case. In this ideal case\textbf{,} with the perfect pulse-duration and
frequency-matching conditions, we can prepare the perfect target state.
However, in practical cases, the system cannot avoid environmental effects.
Moreover, the imperfection of the parameters chosen also affects the fidelity
of the target state. For example, different $N$ describe different Bessel
functions for effective coupling strengths between the cavity field, the
two-level system\textbf{,} and the classical driving field. Then the
optimization for the target state will also be different. For concreteness, as
an example, let us study the effects of both the environment and imperfect
parameters on the target state
\begin{equation}
\left\vert \psi_{02}\right\rangle =\frac{1}{\sqrt{2}}\left(  \left\vert
0\right\rangle +\left\vert 2\right\rangle \right)  \left\vert g\right\rangle ,
\label{eq:37}%
\end{equation}
in the displacement picture, whose density matrix operator can be given as
\begin{equation}
\rho^{I}=\left\vert \psi_{02}\right\rangle \left\langle \psi_{02}\right\vert .
\end{equation}
We also assume that the terms with the Bessel function for $N=-1$ are chosen
for the state preparation. But other terms with the Bessel function order
$N^{\prime}\neq N$ are also involved. Thus we have to choose $\omega_{z}%
\neq\omega n/\left(  N^{\prime}-N\right)  $ to minimize the effect of these
terms. Among all the terms with the Bessel function order $N^{\prime}$, the
dominant ones are those with $N^{\prime}=0,1,\pm2$. That is, the chosen
$\omega_{z}$ has to satisfy the condition $\omega_{z}\neq$ $n\omega
,\,n\omega/2$, and $n\omega/3$.

To study the environmental effect on the state preparation, we assume that the
dynamical evolution of the system satisfies the following master equation
\begin{equation}
\dot{\rho}=-i\left[  H,\rho\right]  +\mathcal{L}_{q}\left[  \rho\right]
+\mathcal{L}_{c}\left[  \rho\right]  , \label{eq:ME}%
\end{equation}
when the environmental effect is taken into account, where the Hamiltonian $H$
is given by Eq.~(\ref{eq:1}) and
\begin{align}
\mathcal{L}_{q}\left[  \rho\right]   &  =\frac{1}{2}\sum_{1\geq j\geq i\geq
0}\gamma_{ij}\left(  2\tilde{\sigma}_{ij}\rho\tilde{\sigma}_{ji}-\tilde
{\sigma}_{jj}\rho-\rho\tilde{\sigma}_{jj}\right)  ,\\
\mathcal{L}_{c}\left[  \rho\right]   &  =\frac{\kappa}{2}\left(  2a\rho
a^{\dag}-a^{\dag}a\rho-\rho a^{\dag}a\right)  ,
\end{align}
describe the dissipation of the qubit and the cavity field, respectively. Here
$\rho$ is the reduced density matrix of the qubit and the cavity field. And $\sigma_{ij}=\vert i \rangle \langle j \vert$, where we define $\vert 0 \rangle\equiv\vert g \rangle$, and $\vert 1 \rangle\equiv\vert e \rangle$. The
operators $\tilde{\sigma}_{ij}$ are given by
\[
\tilde{\sigma}_{ij}=R_{y}\left(  \theta\right)  \sigma_{ij}R_{y}^{\dagger
}\left(  \theta\right)  ,
\]
with $R_{y}\left(  \theta\right)  =\exp\left(  -i\theta\sigma_{y}/2\right)  $
and $\theta=\arctan\left(  \omega_{x}/\omega_{z}\right)  $. This is because we
have used the eigenstates of $\sigma_{z}$ as a basis (persistent current
basis) to represent the Hamiltonian of the qubit. Note that $\gamma_{10}$ is
the the relaxation rate, while $\gamma_{11}$ and $\gamma_{00}$ are the
dephasing rates. The parameter $\kappa$ is the decay rate of the cavity field.
In the following calculations, we assume $\gamma_{00}=0$.

We first neglect the environmental effects and just study how the unwanted
terms with the Bessel function for $N^{\prime}\neq-1$ affects the fidelity for
different parameters $x_{d}=2\Omega_{d}/\omega_{d}$ and $\eta=2g/\omega$ of
the driving field and the cavity field when the target state in
Eq.~(\ref{eq:37}) is prepared. We define the density matrix
\begin{equation}
\rho^{D}=\left\vert \psi_{02}^{D}\right\rangle \left\langle \psi_{02}%
^{D}\right\vert =D^{\dag\!\!}\left(  \eta/2\right)  \rho^{I}D^{\!\!}\left(
\eta/2\right)  \label{eq:41}%
\end{equation}
which is the ideal target state in the original picture. The actual target state
generated in the original picture is denoted by the density operator $\rho
^{A}$ when the effect of the unwanted terms is taken into account. The
Fidelity for the target state is then given by
\begin{equation}
F=\operatorname{Tr}\left\{  \rho^{A}\rho^{D}\right\}  .
\end{equation}

Let us now take the parameters $\omega/2\pi=2$ GHz and $\omega_{z}/2\pi=19.5$
GHz as an example to show how the parameters affect the fidelity. The highly
symmetry-broken condition is satisfied by taking, e.g., $\omega_{x}/2\pi=1.6$
GHz. We have listed the fidelities for different $\eta$ and $x_{d}$ in
Table~\ref{tab-II}, from which we can find that larger values\textbf{ }of
$\eta$ and $x_{d}$ are more likely to induce a higher fidelity. Because in the
range considered for $\eta$ and $x_{d}$, a larger $x_{d}$ can enhance the
desired term through making $J_{-1}\left(  x_{d}\right)  $ larger [see
Fig.~\ref{fig2}(a)] while $\eta$ achieves the same goal by enhancing the Rabi
frequency for $\left\vert 0\right\rangle \left\vert e\right\rangle
\longleftrightarrow$ $\left\vert 2\right\rangle \left\vert g\right\rangle $
(see Fig.~\ref{fig7} in Appendix \ref{Append:Opt1}).

In Table~\ref{tab-II}, the largest fidelity is $F_{m}=0.886$ which occurs at
the optimal parameters $\eta=\eta_{m}=1.11$, and $x_{d}=x_{d}^{m}=$ $1.305$,
where we have also obtained the total time $T_{m}=1.82$ ns for generating
the target state. From the above numerical calculations, we show that the
fidelity of the prepared target state is significantly affected by the
parameters of the qubit, cavity field, and driving field.
Note that the fidelities in Table~\ref{tab-II} may not be satisfactory
for practical applications in quantum information processing, which may require fidelities approaching 100\%.  However, the fidelity can be further optimized by carefully
choosing suitable experimental parameters. For instance, $\eta=1.5$ and
$x_{d}=1.305$ would produce a more desirable fidelity of
0.9143, and it is still possible to obtain much higher fidelities
when related parameters are further optimized. We should also mention that the
effect of the unwanted terms can be totally avoided if for each generation
step, the control pulses for the driving frequency $\omega_{d}$, driving
strength $\Omega_{d}$, driving phase $\phi_{d}$, and the pulse duration $t$ are
all perfectly designed to compensate the effect of the unwanted terms.%

\begin{table}[tbp] \centering
\caption{The fidelities of the target state are listed for different values of the parameters $x_d=2\Omega_d/\omega_d$ and $\eta=2g/\omega$. Here we have chosen $\omega_{z}/2\pi=19.5\operatorname{GHz}$, $\omega_{x}=1.6\operatorname{GHz}$, and $\omega=2\operatorname{GHz}$.}\qquad%
\begin{tabular}
[c]{c|c|ccccc}\hline\hline
\multicolumn{2}{c|}{} & \multicolumn{5}{|c}{Lamb-Dicke parameter
$\eta=2g/\omega$}\\\cline{3-7}%
\multicolumn{2}{c|}{} & $0.330$ & $0.590$ & $0.850$ & $1.110$ & $1.370$%
\\\hline
& $0.265$ & $0.240$ & $0.219$ & $0.237$ & $0.347$ & $0.332$\\
& $0.525$ & $0.281$ & $0.282$ & $0.621$ & $0.719$ & $0.734$\\
$x_{d}=2\Omega_{d}/\omega$ & $0.785$ & $0.445$ & $0.416$ & $0.722$ & $0.827$ &
$0.841$\\
& $1.045$ & $0.369$ & $0.519$ & $0.780$ & $0.857$ & $0.877$\\
& $1.305$ & $0.335$ & $0.530$ & $0.791$ & $0.886$ & $0.879$\\\hline\hline
\end{tabular}
\label{tab-II}%
\end{table}%

Now we study the environmental effect on the fidelity of the prepared state by
taking experimentally achievable parameters, e.g., $\gamma_{10}/2\pi
=\kappa/2\pi=1$ MHz and $\gamma_{11}=2$ MHz. We also choose $\eta=\eta_{m}$
and $x_{d}=x_{d}^{m}$, and other parameters (i.e., $\omega_{x}$, $\omega_{z}$,
and $\omega$) are kept the same as in Table \ref{tab-II}. Now the fidelity we
obtain via numerical calculations is $F_{m}^{\prime}=0.8775$.

The Wigner function represents the full information of the states of the
cavity field and can be measured via quantum state
tomography~\cite{Leibfried1996}. The Wigner function of the cavity field has
recently been measured in circuit QED systems~\cite{Eichler2011,Shalibo2013}.
To obtain the state of the cavity field, let us now trace out the qubit part
of the density operator for the qubit-cavity composite system using
the formula
\begin{equation}
\rho_{c}^{p}=\operatorname{Tr}_{q}\rho^{p}=\left\langle g\right\vert \rho
^{p}\left\vert g\right\rangle +\left\langle e\right\vert \rho^{p}\left\vert
e\right\rangle , \label{eq:trace_rho}%
\end{equation}
where $p$ refers to either $D$, $I$, or $A$. Here, $\rho^{I}$ is the ideal
target state in the displacement picture, $\rho^{D}$ is the the ideal target
state in the original picture, and $\rho^{A}$ is the actual target state in the
original picture. Therefore, $\rho_{c}^{p}$ is the cavity part of
the qubit-cavity-composite state $\rho^{p}$.  It should be emphasized here that the actual state
$\rho^{A}$ denotes the generated target state in the original picture with the
same parameters as in the ideal case, but including the effects of both the
environment and unwanted terms. By definition, given an arbitrary
density operator $\rho,$ the Wigner function $\mathcal{W}\left(  \beta
,\beta^{\ast}\right)  $ and the Wigner characteristic function $C^{W}\left(
\lambda,\lambda^{\ast}\right)  $ have the following
relations~\cite{Agarwal1991,Louisell1973,QuantumNoise},%
\begin{align}
C^{W}\left(  \lambda,\lambda^{\ast}\right)   &  =\operatorname{Tr}\left\{
\rho\exp\left(  \lambda a^{\dag}-\lambda^{\ast}a\right)  \right\}
,\label{eq:CW}\\
\mathcal{W}\left(  \beta,\beta^{\ast}\right)   &  =\int\frac{\text{\textrm{d}%
}^{2}\lambda}{\pi^{2}}\;C^{W}\left(  \lambda,\lambda^{\ast}\right)
\exp\left(  -\lambda\beta^{\ast}+\lambda^{\ast}\beta\right)  .
\label{eq:Wigner}%
\end{align}
Moreover, if $\rho$ is expanded in the Fock state space, i.e,
\begin{equation}
\rho=\sum_{mn}\rho_{mn}\left\vert m\right\rangle \left\langle n\right\vert ,
\end{equation}
then we have the Wigner function of $\rho$ given by
\begin{equation}
\mathcal{W}\left(  \beta,\beta^{\ast}\right)  =\sum_{mn}\rho_{mn}%
\mathcal{W}_{mn}\left(  \beta,\beta^{\ast}\right)  , \label{eq:Wigner_rho_mn}%
\end{equation}
where%
\begin{equation}
\mathcal{W}_{mn}\left(  \beta,\beta^{\ast}\right)  =\left\{
\begin{array}
[c]{ll}%
\begin{array}
[c]{l}%
\frac{2^{n-m+1}}{\pi}\left(  -1\right)  ^{m}\sqrt{\frac{m!}{n!}}\;\beta
^{n-m}\\
\times e^{-2\left\vert \beta\right\vert ^{2}}L_{m}^{\left(  n-m\right)
}\left(  4\left\vert \beta\right\vert ^{2}\right)  ,
\end{array}
& m<n,\\%
\\
\begin{array}
[c]{l}%
\frac{2^{m-n+1}}{\pi}\left(  -1\right)  ^{n}\sqrt{\frac{n!}{m!}}\;\beta^{\ast
m-n}\\
\times e^{-2\left\vert \beta\right\vert ^{2}}L_{n}^{\left(  m-n\right)
}\left(  4\left\vert \beta\right\vert ^{2}\right)  ,
\end{array}
& m\geq n.
\end{array}
\right.  \label{eq:Wmn}%
\end{equation}

\begin{figure}[ptb]
\includegraphics[bb=165 184 419 642,width=0.48 \textwidth,clip]{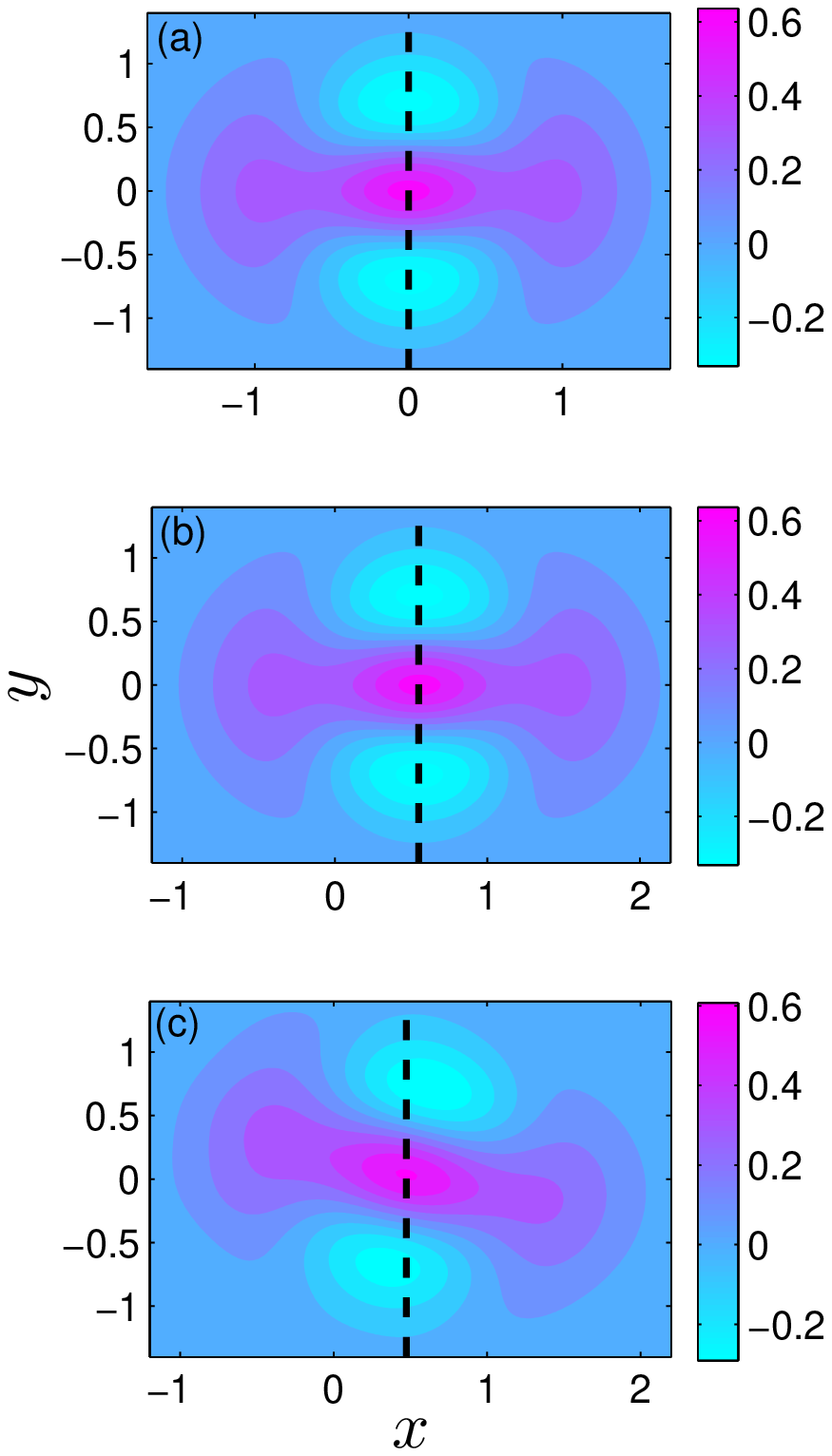}\newline%
\caption{Wigner functions (a) $\mathcal{W}_{c}^{I}\left(  \beta,\beta^{\ast
}\right)  $, (b) $\mathcal{W}_{c}^{D}\left(  \beta,\beta^{\ast}\right)  $, (c)
$\mathcal{W}_{c}^{A}\left(  \beta,\beta^{\ast}\right)  $ respectively for (a)
$\rho_{c}^{I}$, (b) $\rho_{c}^{D}$, and (c) $\rho_{c}^{A}$. Here
$x=\operatorname{Re}\left(  \beta\right)  $, and $y=\operatorname{Im}\left(
\beta\right)  .$ In the above figures, the parameters are chosen
as $\omega_{z}/2\pi=19.5\operatorname{GHz}$, $\omega_{x}=1.6\operatorname{GHz}%
$, $\omega=2\operatorname{GHz}$, $\gamma_{21}/2\pi=\kappa/2\pi
=1\operatorname{MHz}$, $\gamma_{11}=0$, $\gamma_{22}=0.1\operatorname{MHz}$,
$\eta=\eta_{m}=$ $1.11$, and $x_{d}=x_{d}^{m}=$ $1.305$. We have used a
vertical dashed line in (a), (b), and (c) to highlight the displacement of the
central point, also the maximum point, of the Wigner functions. As
shown in (a), the central point of $\mathcal{W}_{c}^{I}\left(  \beta
,\beta^{\ast}\right)  $ is the origin. Since $\rho_{c}%
^{D}=D\!\left(  \eta_{m}/2\right)  \rho_{c}^{I}\:D^{\dagger}\!\left(  \eta
_{m}/2\right)  $, the Wigner function for $\rho_{c}^{D}$, $\mathcal{W}_{c}%
^{D}\left(  \beta,\beta^{\ast}\right)  $ and that for $\rho_{c}^{I}$,
$\mathcal{W}_{c}^{I}\left(  \beta,\beta^{\ast}\right)  $ have the relation
$\mathcal{W}_{c}^{D}\left(  \beta,\beta^{\ast}\right)  =\mathcal{W}_{c}%
^{I}(\beta-\eta_{m}/2,\beta^{\ast}-\eta_{m}/2)$. This is the very reason why
(a) and (b) exhibit the same profile as well as a horizonal translation. The
exact value of this translation length is, of course, $\eta_{m}/2=0.555$.
Compared to the ideal target state in the original picture, i.e.,
$\mathcal{W}_{c}^{D}\left(  \beta,\beta^{\ast}\right)  $ in (b), the actual
target state $\mathcal{W}_{c}^{A}\left(  \beta,\beta^{\ast}\right)  $ in (c)
possesses nearly the same central point. But due to the effects of environment
and unwanted terms, there appears, in (c), a new feature of small local twists and a
global rotation.}%
\label{fig6}%
\end{figure}

As shown in Eq.~(\ref{eq:Wigner_rho_mn}), the Wigner function and the density
operator can in principle be derived from each other, which are closely
related by the function $\mathcal{W}_{mn}\left(  \beta,\beta^{\ast}\right)  $.
If $\left\{  \mathcal{W}_{mn}\left(  \beta,\beta^{\ast}\right)  \right\}  $
are taken as the basis functions, then $\rho_{mn}$ can be considered as the
spectrum of $\mathcal{W}\left(  \beta,\beta^{\ast}\right)  $. Moreover, if we
define $\rho^{D}=D\!\!\left(  \alpha\right)  \rho\:D^{\dag}\!\!\left(
\alpha\right)  $ and its Wigner function as $\mathcal{W}^{D}\left(
\beta,\beta^{\ast}\right)  $, through the definitions in Eq.~(\ref{eq:CW}) and
Eq.~(\ref{eq:Wigner}), we can easily obtain
\begin{equation}
\mathcal{W}^{D}\left(  \beta,\beta^{\ast}\right)  =\mathcal{W}\left(
\beta-\alpha,\beta^{\ast}-\alpha^{\ast}\right)  .
\end{equation}
It is clear that the displacement operator $D\!\left(  \alpha\right)  $
displaces the Wigner function by $\alpha$ in the coordinate system. Since
$\rho_{c}^{D}=D\!\left(  \eta_{m}/2\right)  \rho_{c}^{I}\:D^{\dagger}\!\left(
\eta_{m}/2\right)  $, the Wigner function for $\rho_{c}^{D}$, $\mathcal{W}%
_{c}^{D}(\beta,\beta^{\ast})$ and that for $\rho_{c}^{I}$, $\mathcal{W}%
_{c}^{I}(\beta,\beta^{\ast})$ must have the relation $\mathcal{W}_{c}%
^{D}(\beta,\beta^{\ast})=\mathcal{W}_{c}^{I}(\beta-\eta_{m}/2,\beta^{\ast
}-\eta_{m}/2)$. Therefore, Fig.~\ref{fig6}(a), i.e., the figure for $\mathcal{W}%
_{c}^{I}$ and Fig.~\ref{fig6}(b), i.e., the figure for $\mathcal{W}_{c}^{D}$, are in
fact of the same profile except that there is a horizonal translation between
them. In Figs.~\ref{fig6}(a,b,c), the vertical dashed line that goes through the maximum value of the
Wigner function indicates the horizonal component of its central
position. Since the displacement operator between Fig.~\ref{fig6}%
(a) and Fig.~\ref{fig6}(b) is $D\left(  \eta_{m}/2\right)  $, then the amount
of the translation is $\eta_{m}/2=0.555$. When including the environment and
unwanted terms, Fig.~\ref{fig6}(c) shows how the Wigner function becomes
different from Fig.~\ref{fig6}(b). We can determine that the displacements of
Fig.~\ref{fig6}(b) and Fig.~\ref{fig6}(c) are basically the same. But a
careful comparison shows that the horizonal central position of
Fig.~\ref{fig6}(c) is $0.4745$, $0.0805$, which is less than that
of Fig. \ref{fig6}(b), which is $\eta_{m}/2=0.555$. We think this small
difference can be mainly attributed to the effect of
the environment and unwanted terms. Figure~\ref{fig6}(c) also shows local twists as well as a
global rotation compared with Fig.~\ref{fig6}(b). The global rotation
represents the average phase noise, while the local twists represent the
corresponding fluctuations. Though both the environment and unwanted terms
both affect the fidelity of the states prepared, our calculations show that
under the specified parameters, the role of the unwanted terms is dominant
when the imperfect pulses are applied to state preparation, since the
generation time $T_{m}=1.82$ ns is far from inducing serious decoherence at
the specified decay rates, which is well manifested by the poor fidelity
reduction $F_{m}-F_{m}^{\prime}=0.0085$. Recall that $F_{m}$ is the fidelity obtained using
the optimal paramters in Table.~\ref{tab-II} when only including the effects of
the unwanted terms, while $F_{m}^{\prime}$ is the fidelity obtained using the
same parameters, but with the effects of both the environment and unwanted
terms considered.

\section{Discussions}

Let us now discuss the feasibility of the experiments for the generation of
nonclassical microwave states using superconducting quits interacting with a
single-mode microwave field. The frequency of the qubit cannot be extremely
large. Thus the maximum photon number in multiphoton processes is limited by
the ratio $\omega_{z}/\omega$\textbf{, }between the frequency $\omega_{z}$ of
the qubit and that of the cavity field $\omega$. This means that the qubits
should be far away from the optimal point for the flux and charge qubits when
the microwave states are generated using our proposed methods. This might be a
problem for the preparation of arbitrary superpositions, because the coherence
time becomes short when the flux or charge qubit deviates from the optimal
point. However, for the particular number state $\left\vert n\right\rangle $,
there is no requirement for the coherence and thus it should be more
efficient, because we need only to prepare the qubit in the excited state, and
then the state $\left\vert n\right\rangle $ can be prepared via an $n$-photon
red-sideband excitation. We know that the phase~\cite{martinis1,martinis2} and
Xmon~\cite{Xmon} qubits are not very sensitive to the optimal point. Thus the
proposal might be more efficient for these qubis coupled to a microwave
cavity. It should be noted that the imperfect pulse can significantly affect
the fidelity. We thus suggest that enough optimization be implemented to reach
an acceptable fidelity.

\section{conclusions}

We have proposed a method to prepare nonclassical microwave states via
longitudinal-coupling-induced multi-photon processes when a driven
symmetry-broken superconducting qubit is coupled to a single-mode microwave
field. With controllable $k$-photon processes in a SQ with a
symmetry-broken potential energy,
only $n_{\max}+1$ steps are needed to synthesize the superposition of Fock
states with the largest photon number $n_{\max}$.
However, in contrast to the method used in Refs.~\cite{liuepl,martinis2},
 with one-photon processes in the SQ inside the cavity,
$2n_{\mathrm{max}}$ steps are needed to synthesize the same state. Moreover, using
$k$-photon processes, a $k$-photon Fock state $\left\vert k\right\rangle $ can
be generated with just two steps, while with one-photon processes,  $2k$
steps are required to produce the same state. Thus, the
time to generate the same state using multiphoton processes is shorter than that using only a single-photon process. Therefore the fidelity should also be improved.
In this sense, our method is more efficient than the one
in Refs.~\cite{liuepl,martinis2}. Besides, we have provided an
analytical solution for the total time needed to generate a target state
$\left\vert \psi_{n_{\max}}\right\rangle $.

We have made a detailed analysis of the ground state when the system is
sufficiently cooled. We find that in the highly-symmetry-broken and
strong-coupling case, the ground state can still be regarded as the vacuum
state in the displacement picture. The displacement effect on both the initial
state and the target state has also been studied. Generally, the displacement
will induce a variation of the photon-number distribution. But in the
representation of the Wigner function, its influence is just a shift of the
center of the Wigner function by the Lamb-Dicke parameter\textbf{ }%
$\eta=2g/\omega$ between the coupling strength $g$ of the cavity field to the
qubit and the frequency\textbf{ }$\omega$ of the cavity field. We note that
the Fock state produced in the displacement picture is a displaced number
state in the original picture. Thus, a circuit QED system with broken
symmetry in the qubit potential energy can be used to easily generate a
displaced number state. This can be used to study the boundary between the
classical and quantum
worlds~\cite{Fink,Fedorov,NoriQuanClass2008,NoriQCPRB2008}.

In summary, although we find that the nonclassical photon state can be more
easily produced when the symmetry of the potential energy of the SQ is broken,
this method can be applied to any device with longitudinal and transverse
couplings to two-level systems. Although the Fock state can be produced in any
symmetry-broken qubit, the superposition of Fock states might be easily
realizable in a circuit QED system formed by a phase qubit and a cavity field.
This is because phase qubits have no optimal point, and thus not
sensitive to the working point of the external parameter. Our proposal is
experimentally realizable with current technology.

\section{Acknowledgements}

YXL is supported by the National Basic Research Program of China Grant
No.~2014CB921401, the NSFC Grants No.~61025022, and No.~91321208. FN is
partially supported by the RIKEN iTHES Project, MURI Center for Dynamic
Magneto-Optics, and a Grant-in-Aid for Scientific Research (S).

\appendix

\section{Detailed steps for generating the nonclassical
state\label{Append:Opt1}}

If we substitute Eq.~(\ref{eq:Ur}), Eq.~(\ref{eq:Uc}), and
Eq.~(\ref{eq:psi_n_expand}) into Eq.~(\ref{eq:psi_recur}), then the following
relations can be obtained, e.g., for the generation of $\left\vert \psi
_{0}\right\rangle $, i.e., for the step $n=0$,
\begin{align}
C_{0g}^{\left(  0\right)  }  &  =\exp\left(  i\alpha_{0g}^{\left(  0\right)
}\right)  \cos\left(  \left\vert \Omega_{N,c}^{0,0}\right\vert t_{0}\right)
C_{0g}^{\left(  -1\right)  },\label{eq:recur0}\\
C_{0e}^{\left(  0\right)  }  &  =\exp\left(  i\alpha_{0e}^{\left(  0\right)
}\right)  \mathrm{sin}\left(  \left\vert \Omega_{N,c}^{0,0}\right\vert
t_{0}\right)  C_{0g}^{\left(  -1\right)  }, \label{eq:recur0_1}%
\end{align}
with $|C_{0g}^{\left(  -1\right)  }|=1$\textbf{, }which is determined by the
initial condition. However, for the generation of $\left\vert \psi
_{n}\right\rangle $ with $n\geq1$ from the state $\left\vert \psi
_{n-1}\right\rangle $, we can obtain the following relations for their
coefficients
\begin{align}
C_{kg}^{(n)}  &  =\exp\left(  i\alpha_{kg}^{n}\right)  C_{kg}^{\left(
n-1\right)  },\label{eq:recur1}\\
C_{ng}^{(n)}  &  =\exp\left(  i\alpha_{ng}^{n}\right)  \mathrm{sin}\left(
\left\vert \Omega_{N,r}^{n,0}\right\vert t_{n}\right)  C_{0e}^{\left(
n-1\right)  },\label{eq:recur2}\\
C_{0e}^{(n)}  &  =\exp\left(  i\alpha_{0e}^{n}\right)  \cos\left(  \left\vert
\Omega_{N,r}^{n,0}\right\vert t_{n}\right)  C_{0e}^{\left(  n-1\right)  },
\label{eq:recur3}%
\end{align}
with $k\leq n-1$. Here, the phases $\alpha_{0g}^{(0)}$ and $\alpha_{0e}^{(0)}$
for $n=0$ are determined by
\begin{align}
\alpha_{0g}^{\left(  0\right)  }  &  =\frac{x_{d}}{2}\mathrm{sin}\left(
\omega_{d}^{\left(  0\right)  }t_{0}+\phi_{d}^{\left(  0\right)  }\right)
-\frac{x_{d}}{2}\mathrm{sin}\left(  \phi_{d}^{\left(  0\right)  }\right)
+\frac{\omega_{z}t_{0}}{2},\\
\alpha_{0e}^{\left(  0\right)  }  &  =-\frac{x_{d}}{2}\mathrm{sin}\left(
\omega_{d}^{\left(  0\right)  }t_{0}+\phi_{d}^{\left(  0\right)  }\right)
-\frac{x_{d}}{2}\mathrm{sin}\left(  \phi_{d}^{\left(  0\right)  }\right)
-\frac{\omega_{z}t_{0}}{2}\nonumber\\
&  +\phi_{0}-\frac{\pi}{2}.
\end{align}
The other phases with for $n\geq1$ are given by
\begin{align}
\alpha_{kg}^{\left(  n\right)  }  &  =\frac{x_{d}}{2}\mathrm{sin}\left(
\omega_{d}^{\left(  n\right)  }t_{n}+\phi_{d}^{\left(  n\right)  }\right)
-\frac{x_{d}}{2}\mathrm{sin}\left(  \phi_{d}^{\left(  n\right)  }\right)
+\frac{\omega_{z}t_{n}}{2}\nonumber\\
&  -k\omega t_{n},\\
\alpha_{ng}^{\left(  n\right)  }  &  =\frac{x_{d}}{2}\mathrm{sin}\left(
\omega_{d}^{\left(  n\right)  }t_{n}+\phi_{d}^{\left(  n\right)  }\right)
+\frac{x_{d}}{2}\mathrm{sin}\left(  \phi_{d}^{\left(  n\right)  }\right)
+\frac{\omega_{z}t_{n}}{2}\nonumber\\
&  -n\omega t_{n}-\phi_{n}-\frac{\pi}{2},\\
\alpha_{0e}^{\left(  n\right)  }  &  =-\frac{x_{d}}{2}\mathrm{sin}\left(
\omega_{d}^{\left(  n\right)  }t_{n}+\phi_{d}^{\left(  n\right)  }\right)
+\frac{x_{d}}{2}\mathrm{sin}\left(  \phi_{d}^{\left(  n\right)  }\right)
-\frac{\omega_{z}t_{n}}{2},
\end{align}
where $k\leq n-1.$ In Eqs. (\ref{eq:recur0}-\ref{eq:recur3}),
\begin{equation}
\phi_{n}=\left\{
\begin{array}
[c]{ll}%
\arg\left(  \Omega_{N,c}^{0,0}\right)  =N\phi_{d}^{\left(  n\right)  }-\pi, &
\ \ \ n=0,\\
\arg\left(  \Omega_{N,r}^{n,0}\right)  =N\phi_{d}^{\left(  n\right)  }-\left(
n+1\right)  \pi, & \ \ \ n\geq1,
\end{array}
\right.
\end{equation}
if we select an $N$ and $x_{d}$ such that $J_{N}\left(  x_{d}\right)
<0$, and
\begin{equation}
\phi_{n}=\left\{
\begin{array}
[c]{ll}%
\arg\left(  \Omega_{N,c}^{0,0}\right)  =N\phi_{d}^{\left(  n\right)  }, &
\ \ \ n=0,\\
\arg\left(  \Omega_{N,r}^{n,0}\right)  =N\phi_{d}^{\left(  n\right)  }-n\pi, &
\ \ \ n\geq1,
\end{array}
\right.
\end{equation}
if we select an $N$ and $x_{d}$ such that $J_{N}\left(  x_{d}\right)
>0$. Here, $\omega_{d}^{\left(  n\right)  },$ $\phi_{d}^{\left(
n\right)  },$ and $t_{n}$ are, respectively, the driving frequency,
driving phase, and time duration for each generation step. From
Eq.~(\ref{eq:recur1}), we know that%
\begin{equation}
\left\vert C_{k0}^{\left(  n\right)  }\right\vert =\left\vert C_{k0}^{\left(
n_{\max}\right)  }\right\vert =\left\vert C_{k}\right\vert ,\ \ \ k\leq n.
\end{equation}
and hence%
\begin{equation}
\left\vert C_{0e}^{\left(  n-1\right)  }\right\vert =\left(  1-\sum
_{k=0}^{n-1}\left\vert C_{kg}^{\left(  n-1\right)  }\right\vert ^{2}\right)
^{1/2}=\left(  \sum_{k=n}^{n_{\max}}\left\vert C_{k}\right\vert ^{2}\right)
^{1/2}.
\end{equation}
Then, from Eqs.~(\ref{eq:recur0}, \ref{eq:recur0_1}) and Eqs. (\ref{eq:recur2},
\ref{eq:recur3}), we respectively have%
\begin{align}
\left\vert \Omega_{N,c}^{0,0}\right\vert t_{0}  &  =\arccos\left\vert
\frac{C_{0g}^{\left(  0\right)  }}{C_{0g}^{\left(  -1\right)  }}\right\vert
+2l\pi\nonumber\\
&  =\arccos\left\vert C_{0g}^{\left(  0\right)  }\right\vert +2l\pi ,
\label{eq:t0}\\
\left\vert \Omega_{N,r}^{n,0}\right\vert t_{n}  &  =\arcsin\left( \frac{\left\vert
C_{ng}^{\left(  n\right)  }\right\vert}{\left\vert C_{0e}^{(n-1)}\right\vert}
\right)  +2l\pi\nonumber\\
&  =\arcsin\left[  \frac{\left\vert C_{n}\right\vert}{\left(  \sum_{k=n}^{n_{\max}%
}\left\vert C_{k}\right\vert ^{2}\right) ^{1/2}}\right]  +2l\pi. \label{eq:tn}%
\end{align}
where $l$ is an arbitrary integer. Using Eqs.~(\ref{eq:recur0},
\ref{eq:recur0_1}) and Eqs.~(\ref{eq:recur2}, \ref{eq:recur3}), we derive%
\begin{align}
\arg\left(  C_{0g}^{\left(  0\right)  }\right)  -\arg\left(  C_{0e}^{\left(
0\right)  }\right)   &  =\alpha_{0g}^{\left(  0\right)  }-\alpha_{0e}^{\left(
0\right)  }+2l\pi,\\
\arg\left(  C_{ng}^{\left(  n\right)  }\right)  -\arg\left(  C_{0e}^{\left(
n\right)  }\right)   &  =\alpha_{ng}^{\left(  n\right)  }-\alpha_{0e}^{\left(
n\right)  }+2l\pi. \label{eq:arg_nn0_n01}%
\end{align}
If $n=n_{\max}$ in Eq.~(\ref{eq:arg_nn0_n01}), then $C_{0e}%
^{\left(  n\right)  }=0$, with no definition of the phase angle. Thus we can assume that $\arg\left(  C_{0e}^{\left(  n\right)  }\right)  =0$, without
affecting the final result. Here,\textbf{ }$C_{ng}^{\left(
n\right)  }$ and $C_{0e}^{\left(  n\right)  }$ can be obtained through the
following recursion relations
\begin{align}
C_{0e}^{\left(  n-1\right)  }  &  =\left\{
\begin{array}
[c]{ll}%
\frac{C_{0e}^{\left(  n\right)  }}{\exp\left(  i\alpha_{0e}^{\left(  n\right)
}\right)  \cos\left(  \left\vert \Omega_{N,r}^{n,0}\right\vert t_{n}\right)
}, & 1\leq n<n_{\max},\\
\\
\frac{C_{ng}^{\left(  n\right)  }}{\exp\left(  i\alpha_{ng}^{\left(  n\right)
}\right)  \mathrm{sin}\left(  \left\vert \Omega_{N,r}^{n,0}\right\vert
t_{n}\right)  }, & n=n_{\max},
\end{array}
\right. \label{eq:Cn_101}\\
C_{kg}^{\left(  n-1\right)  }  &  =\frac{C_{kg}^{\left(  n\right)  }}%
{\exp\left(  i\alpha_{kg}^{\left(  n\right)  }\right)  },k\leq n-1,\\
C_{0g}^{\left(  -1\right)  }  &  =\frac{C_{0e}^{\left(  0\right)  }}%
{\exp\left(  i\alpha_{0e}^{\left(  0\right)  }\right)  \mathrm{sin}\left(
\left\vert \Omega_{N,c}^{0,0}\right\vert t_{0}\right)  } .\label{eq:C_100}%
\end{align}
In Eq.~(\ref{eq:Cn_101}), distinguishing the case when $n=n_{\max}$ from the
other ones is needed to avoid the apperance of $0/0.$
Though Eq.~(\ref{eq:C_100}) implies that $C_{0g}^{\left(  -1\right)  }$ may
have a definite phase, such a phase could only add a global phase factor to
the target state. So it is convenient to directly specify $C_{0g}^{\left(
-1\right)  }=1$.

\begin{figure}[ptb]
\includegraphics[width=0.48\textwidth, clip]{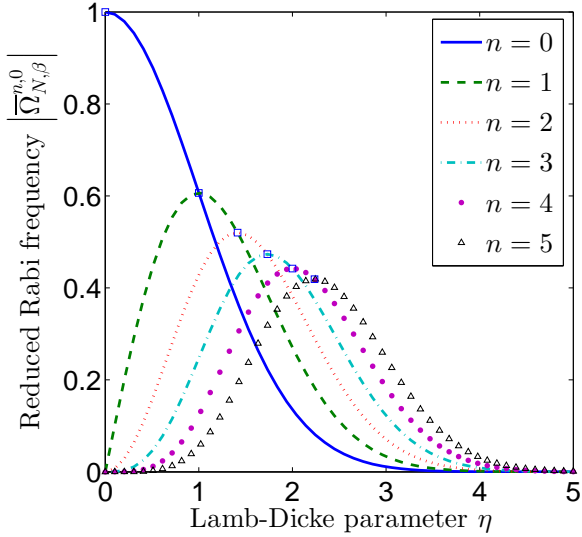}\caption{(color online)
Reduced Rabi frequency $\left\vert \bar{\Omega
}_{N,\beta}^{n,0}\right\vert $, from Eq.~(\ref{eq:Omega_reduced}), as a function
of $\eta$, for $n=1,2,3,4$, and $5$, respectively. The square on each plot
denotes the point that achieves the largest $\left\vert \bar{\Omega}_{N,\beta
}^{n,0}\right\vert $. Recall that $\eta=2g/\omega$, where $g$ is the
qubit-cavity coupling constant and $\omega$ is the frequency of the
single-mode cavity field. Thus, $\eta$ is the normalized coupling. }%
\label{fig7}%
\end{figure}

\begin{figure}[ptb]
\includegraphics[width=0.48\textwidth, clip]{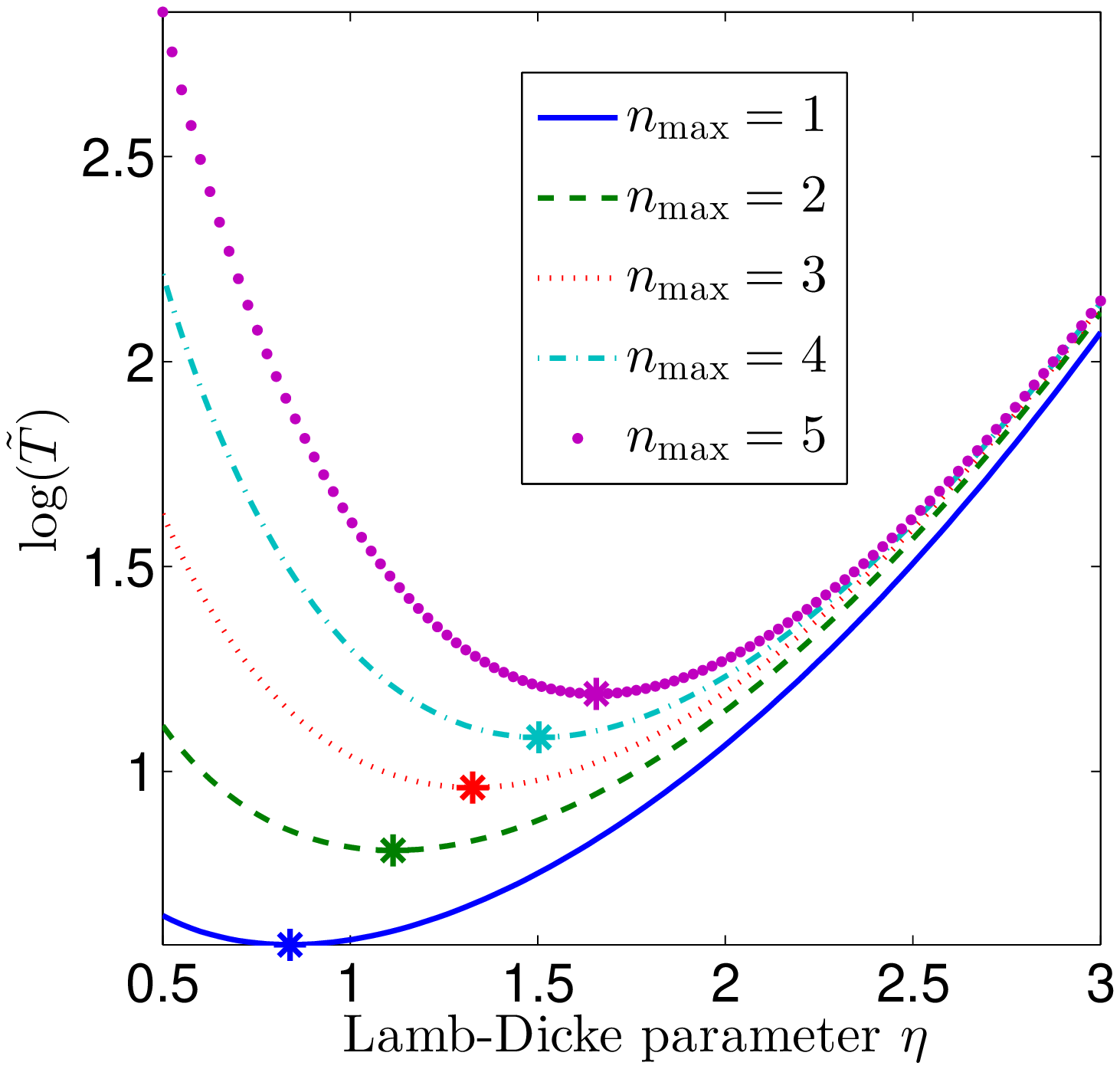}\caption{(color online)
Plot of $\log\left(  \tilde{T}\right)  $ as a function of the Lamb-Dicke
parameter $\eta=2g/\omega$ for $\left\vert \psi_{n_{\max}}\right\rangle
=\sum_{n=0}^{n_{\max}}\left\vert n\right\rangle \bigotimes\left\vert
0\right\rangle /\sqrt{n_{\max}+1}$, with different maximum photon number
$n_{\max}$. Recall that $\tilde{T}=T\left\vert \omega J_{N}\right\vert /2$ is
a normalized time and $T$ is the total time to generate the desired target
state. The star on each curve shows the optimal point where the
normalized generation time reaches its minimum. }%
\label{fig8}%
\end{figure}

We define the reduced Rabi frequency as%
\begin{equation}
\left\vert \bar{\Omega}_{N,\beta}^{n,0}\right\vert =\frac{2\left\vert \Omega
_{N,\beta}^{n,0}\right\vert }{\left\vert \omega_{x}J_{N}\right\vert}=\exp\left(  -\frac{1}{2}\eta^{2}\right)  \frac{\eta^{n}}{\sqrt{n!}}
\label{eq:Omega_reduced}%
\end{equation}
in order to study its dependence on $\eta$. From Eq.~(\ref{eq:Omega_reduced}),
we can obtain the optimal Lamb-Dicke parameter%
\begin{equation}
\eta_{n,\mathrm{o}}=\sqrt{n},
\end{equation}
that achieves the largest reduced Rabi frequency
\begin{equation}
\left\vert \bar{\Omega}_{N,\beta,\mathrm{o}}^{n,0}\right\vert =\exp\left(
-\frac{n}{2}\right)  \frac{n^{n/2}}{\sqrt{n!}},
\end{equation}
which is also the point that makes $\left\vert \bar{\Omega}_{N,\beta}%
^{0,n}\right\vert =\left\vert \bar{\Omega}_{N,\beta}^{0,n-1}\right\vert
$\textbf{, }as illustrated in Fig.~\ref{fig7}. We can also verify
\begin{equation}
\lim_{n\rightarrow\infty}\frac{\left\vert \bar{\Omega}_{N,\beta,\mathrm{o}%
}^{n+1,0}\right\vert }{\left\vert \bar{\Omega}_{N,\beta,\mathrm{o}}%
^{n,0}\right\vert }=\lim_{n\rightarrow\infty}\sqrt{\frac{1}{e}\left(
\frac{n+1}{n}\right)  ^{n}}=1,
\end{equation}
with $\lim_{n\rightarrow\infty}\left\{  \eta_{n+1,\mathrm{o}}/\eta
_{n,\mathrm{o}}\right\}  =1$. This means that when the photon number $n$
increases, the optimal points for the Rabi frequencies between the zero-photon
state and different $n$ photon states tend to approach each other
infinitesimally. But for low photon numbers the optimal points are still
distinguishable from each other.

Let us calculate the total time $T$ for generating the target state%
\begin{align}
T  &  =\sum_{n=0}^{n_{\max}}t_{n}=\frac{2}{\left\vert \omega_{x}%
J_{N}\right\vert }\arccos\left(  \left\vert C_{0}\right\vert \right)
\exp\left(  \frac{\eta^{2}}{2}\right) \nonumber\\
&  +\sum_{n=1}^{n_{\max}}\frac{2\sqrt{n!}}{\left\vert \omega_{x}%
J_{N}\right\vert \eta^{n}}\arcsin\left(  \frac{\left\vert C_{n}\right\vert
}{\sqrt{\sum_{k=n}^{n_{\max}}\left\vert C_{k}\right\vert ^{2}}}\right)
\exp\left(  \frac{\eta^{2}}{2}\right)  ,
\end{align}
where $\eta=2g/\omega$ is the Lamb-Dicke parameter and
we have omitted excessive cycle periods for each step. By taking the
derivative of $T$ with respect to $\eta,$ we can find all the extreme points
of $\eta$, which satisfy the following equation%
\begin{equation}
\sum_{n=-1}^{n_{\max}+1}A_{n}\eta^{n_{\max}+1-n}=0.
\end{equation}
The coefficient $A_{n}$ has been given in Eqs. (\ref{eq:AnBegin}%
)-(\ref{eq:AnEnd}). Further selection among these extreme points and the
experimentally-constrained boundaries of $\eta$ can yield the optimal
Lamb-Dicke parameter $\eta_{\mathrm{opt}}$\textbf{,
}which will lead to the least generation time $T_{\mathrm{opt}}$. Once
$T_{\mathrm{opt}}$ is reached, in principle, the influence of the environment
on the target state fidelity will be minimized.

Similarly to $\left\vert \bar{\Omega}_{N,\beta}^{0,n}\right\vert $, we define%
\begin{equation}
\tilde{T}\left(  \eta\right)  =T\left\vert \omega_{x}J_{N}\right\vert /2.
\end{equation}
The curves of $\log\left(  \tilde{T}\left(  \eta\right)  \right)  $
which for particular states have been plotted in Fig.~\ref{fig8}
with a star on each curve to label the point where the generation time reaches
its least value.

The normalized time needed to generate a target state
is
\begin{align}
\tilde{T}  &  =\sum_{n=0}^{n_{\max}}\tilde{t}_{n}=\arccos\left(  \left\vert
C_{0}\right\vert \right)  \exp\left(  \eta^{2}/2\right) \nonumber\\
&  +\sum_{n=1}^{n_{\max}}\sqrt{n!}\arcsin\left(  \frac{\left\vert
C_{n}\right\vert }{\sqrt{\sum_{k=n}^{n_{\max}}\left\vert C_{k}\right\vert
^{2}}}\right)  \frac{e^{\eta^{2}/2}}{\eta^{n}},
\end{align}
whose extreme points still satisfy%
\begin{equation}
\sum_{n=-1}^{n_{\max}+1}A_{n}\:\eta^{n_{\max}+1-n}=0,
\end{equation}
for unbound $\eta$\textbf{,} where if $n_{\max}=0$,%
\begin{equation}
A_{n}=\left\{
\begin{array}
[c]{ll}%
\arccos\left(  \left\vert C_{0}\right\vert \right)  , & n=-1,\\
0, & \mathrm{others,}%
\end{array}
\right.  \label{eq:AnBegin}%
\end{equation}
if $n_{\max}=1,$%
\begin{equation}
A_{n}=\left\{
\begin{array}
[c]{ll}%
\arccos\left(  \left\vert C_{0}\right\vert \right)  , & n=-1,\\
P_{n+1}, & n=0,\\
-\left(  n-1\right)  P_{n-1}, & n=2,\\
0, & \mathrm{others,}%
\end{array}
\right.
\end{equation}
if $n_{\max}=2$,
\begin{equation}
A_{n}=\left\{
\begin{array}
[c]{ll}%
\arccos\left(  \left\vert C_{0}\right\vert \right)  , & n=-1,\\
P_{n+1}, & 0\leq n\leq1,\\
-\left(  n-1\right)  P_{n-1}, & 2\leq n\leq3,\\
0, & \mathrm{others,}%
\end{array}
\right.
\end{equation}
and for other cases, we have
\begin{equation}
A_{n}=\left\{
\begin{array}
[c]{ll}%
\arccos\left(  \left\vert C_{0}\right\vert \right)  , & n=-1,\\
P_{n+1}, & 0\leq n\leq1,\\
P_{n+1}-\left(  n-1\right)  P_{n-1}, & 2\leq n\leq n_{\max}-1,\\
-\left(  n-1\right)  P_{n-1}, & n_{\max}\leq n\leq n_{\max}+1,\\
0, & \mathrm{others.}%
\end{array}
\right.  \label{eq:AnEnd}%
\end{equation}
Here we have used the abbreviation
\begin{equation}
P_{n}=\sqrt{n!}\arcsin\left(  \frac{\left\vert C_{n}\right\vert }{\sqrt
{\sum_{k=n}^{n_{\max}}\left\vert C_{k}\right\vert ^{2}}}\right)  .
\end{equation}

\hspace{5cm}

\end{document}